\documentclass[twocolumn]{aastex62}
\usepackage{amsmath,amstext}
\usepackage[T1]{fontenc}
\usepackage{txfonts} 
\usepackage[figure,figure*]{hypcap} 
\usepackage{chngcntr}

\shorttitle{Counterrotating low-mass galaxies}
\shortauthors{Starkenburg et al.}

\begin{document}

\title{On the origin of star--gas counterrotation in low-mass galaxies}

\author{Tjitske~K.~Starkenburg}
\altaffiliation{tstarkenburg@flatironinstitute.org}
\affil{Flatiron Institute, 162 5th Avenue, New York NY 10010, USA}
\author{Laura.~V.~Sales}
\altaffiliation{Hellman Fellow}
\affil{University of California Riverside, Riverside CA, USA}
\author{Shy~Genel}
\affil{Flatiron Institute, 162 5th Avenue, New York NY 10010, USA}
\affil{Columbia Astrophysics Laboratory, Columbia University, 550 West 120th Street, New York, NY 10027, USA}
\author{Christina~Manzano-King}
\affil{University of California Riverside, Riverside CA, USA}
\author{Gabriela~Canalizo}
\affil{University of California Riverside, Riverside CA, USA}
\author{Lars~Hernquist}
\affil{Harvard-Smithsonian Center for Astrophysics, 60 Garden Street, Cambridge, MA 02138, USA}

\begin{abstract}
Stars in galaxies form from the cold rotationally supported gaseous disks that 
settle at the center of dark matter halos. In the simplest models, 
such angular momentum is acquired early on at the time of collapse 
of the halo and preserved thereafter, implying a well-aligned spin for the stellar 
and gaseous component. Observations however have shown
the presence of gaseous disks in counterrotation with the stars. We
use the Illustris numerical simulations to study the origin of such 
counterrotation in low mass galaxies 
($M_\star = 2 \times 10^9$ - $5 \times 10^{10}\; \rm M_\odot$),
a sample where mergers have not played a significant role. 
Only ${\sim}1\%$ of our sample shows a counterrotating gaseous disk at $z=0$. 
These counterrotating disks arise in galaxies that have had a 
significant episode of gas removal followed by the 
acquisition of new gas with misaligned angular momentum.
In our simulations, we identify two main channels responsible for the gas loss: 
a strong feedback burst and gas stripping during a fly-by passage
through a more massive group environment. Once settled, counterrotation 
can be long-lived with several galaxies in our sample displaying misaligned 
components  consistently for more than $2$ Gyr. As a result, no major correlation 
with the present day environment or structural properties might remain, 
except for a slight preference for early type morphologies and a 
lower than average gas content at a given stellar mass.
\end{abstract}

\section{Introduction}
\label{sec:intro}

The spin of galaxies is believed to be related to that of their
surrounding dark matter halos. In $\Lambda$ Cold Dark Matter models,
the acquisition of angular momentum precedes the full gravitational
collapse and is set during the linear regime.  Dark matter halos have a small
but well-defined amount of angular momentum which gets imprinted early
on when the coupling between the inertia tensor of the proto-halo
material and the surrounding tidal field is maximum
\citep{Hoyle1949,Peebles1969,Doroshkevich1970}. At these early times, baryons are
well-mixed with the dark matter and are therefore subjected to similar
torques with the surrounding tidal field, meaning that they will
initially inherit the same angular momentum as the dark matter
counterpart. 

After this, gravitational collapse proceeds by conserving
the angular momentum approximately with a gas component that,
able to cool via radiative processes, sinks further into the potential
well of the dark matter. To maintain similar spins, the gas increases
its tangential velocity to compensate for the smaller radii,
explaining the rotationally supported nature of disks embedded in
otherwise dispersion-supported dark matter halos
\citep{WhiteRees1978, FallEfstathiou1980,Moetal1998}. Stars 
form out of this gas and inherit its dynamical properties. Therefore,
in the absence of significant merger events, co-rotation between gas and
stars is the most natural outcome of structure formation.

Early numerical simulations were able to reproduce the main
predictions from this tidal torque scenario for the origin of the dark
matter halo spins \citep{EfstathiouJones1979, White1984,BarnesEfstathiou1987}  and further confirmation may be
found in the orientations of the angular momentum of nearby disk
galaxies with respect to the surrounding large scale structure
\citep{Navarro2004}. With the advent of more sophisticated hydrodynamical
simulations it also became clear that baryons undergo a much more
complex evolution than previously envisioned \citep{vandenBosch2002,Abadi2003b, Brook2011,Scannapieco2012, Bryan2013, Ubler2014,Dubois2014, Teklu2015, Zavala2016,ZjupaSpringel2017, DeFelippis2017, Jiangetal2018, GarrisonKimmel2018}.

Feedback from stars and black holes was identified as an essential
ingredient to prevent runaway formation of stars in the early stages
of galaxy formation and to produce realistic looking disk-dominated
galaxies in simulations
\citep{Navarro1991,Navarro1997,Eke2000,Scannapieco2009,Governato2010}.
Providing the coupling to the surrounding gas is efficient, energy
from young stars, supernova explosions and accretion disks around
black holes may cause a significant fraction of the gas in galaxies to
be expelled though galactic outflows, signatures of which have been
successfully identified in observations \citep[e.g.][]{Martin2005,Martin2012,Rubin2014,Cheung2016}. These galactic winds
carry with them not only mass but also angular momentum,
causing a redistribution of the initial spin of the baryons and a
potential decoupling from the spin of the dark matter halo \citep[e.g.][]{DeFelippis2017}.

Remarkably, despite this complicated galaxy assembly process and irrespective of the
fact that only a small fraction of the baryons makes it into a
galaxy, current cosmological simulations find that in the case of disk
galaxies the amount of angular momentum retained in the disk is
comparable to that of the dark matter halo
\citep{Sales2010,Lagos2017,Geneletal2015,DeFelippis2017,Sokolowska2017,El-Badry2018},
recovering one of the key assumptions of traditional galaxy formation
models \citep[e.g.][]{Moetal1998}.  Most important, the alignment
between the galactic and the dark matter spin remains within $20$-$45$  
degrees \citep[e.g.][]{Bett2010,ZjupaSpringel2017}. 

Within this framework, the existence of galaxies with counterrotating
components is puzzling. Individual galaxies containing components that
rotate in opposite or highly inclined directions have been studied for
decades \citep[e.g.][]{Ulrich1975, Rubinetal1992,Rubin1994} and have
been found in a wide range of masses and morphologies
\citep{Rix1992,Prada1996, Bertola1996, Vergani2007,Coccato2011,Davis2011,Serra2014,Coccato2015,Krajonovic2015,Katkov2016}. The
advent of Integral Field Spectroscopy surveys has also shed important
light on their overall structure, with a detailed mapping of their
complex kinematics
\citep{Emsellem2007,Barrera2014,Barrera2015,Cappellari2016, Jin2016,
  Bryant2019}.

Misalignments are often linked to an external origin, such as the
accretion of satellites or the cooling of misaligned gas from the halo
\citep{Balcells1990,Hernquist1991,Barnes1996,Roskar2010,vandevoort2015}; all events
related to the highly non-linear regime of galaxy assembly. Different
kinds of misalignments are observed in nature, including misaligned
gas--stellar components, two counterrotating stellar disks,
kinematically decoupled cores in early type galaxies, and polar ring
galaxies. Over the years, idealized and cosmological numerical
simulations of single objects have shown that a variety of mechanisms
can give rise to counterrotating components, including $(i)$ mergers
with very specific initial conditions \citep{PuerariPfenniger2001,Crockeretal2009}, $(ii)$ instabilities and other internal dynamical
evolution within galaxies \citep{EvansCollett1994, DeRijckeetal2004}
and $(iii)$ misaligned smooth gas accretion \citep{ThakarRyden1996,Bekki1998, Brook2008, Aumer2013, Algorryetal2014,Pizzellaetal2004}. These studies, however, pertain to very specific
conditions and the relevance of such processes for the galaxy
population as a whole remains unclear.

Furthermore, the timescales for counterrotating components to survive
is poorly constrained. For elliptical or disk galaxies, the
gravitational pull of a non-spherical potential (given by the stellar
distribution) onto a misaligned gaseous disk will act as a sink of the
perpendicular angular momentum component, re-aligning the orientation
of the gas within the preferred plane of symmetry of the stars
\citep{Hunter1969,Tohline1982}. Idealized theoretical estimates
suggest that the timescales needed for this differential precession to
totally align (e.g. $0^\circ$) or anti-align ($180^\circ$) the gas and
stellar components are rather short at the centers of galaxies,
requiring typically less than $5$ dynamical times $t_{\rm dyn}$
\citep[see Fig.~3 in ][]{Tohline1982, Steiman1988}. But several
factors come into play for such estimates, and in particular, more
flattened gravitational potentials, smaller distances or the inclusion
of self-gravity for massive gas disks, may shorten the estimated
timescales even further \citep{Hunter1969}. With these caveats in mind, there seems
to be consensus to the idea that misaligned gas disks will quickly
settle onto the more stable $0^\circ$ or $180^\circ$ configurations
with the stars, leaving little room in nature for the display of these
spectacular kinematical oddities.

The stability of perfectly anti-aligned stellar-gas disks 
allows enough time for a second generation of stars to be 
born from the young misaligned gas, giving rise to a galaxy 
with two counterrotating stellar disks. Numerical simulations by
\citet{Algorryetal2014} have shown the formation of at least one of
such within the $\Lambda$CDM scenario for a case where the
filamentary accretion of gas changes direction at some point during
the halo formation. Similarly,
\citet{Brook2008,Roskar2010,Snaith2012,vandevoort2015} report the study of
simulated zoom-in galaxies where the misalignment between the stars
and the gaseous disk, once established, is maintained for several
Gyr thanks to the continuous supply of gas from satellites or from the
halo with an inclined angular momentum with respect to the initial
galaxy.

These findings within the cosmological picture of galaxy assembly
highlight the need to include the idea
of a continuous gas supply with misaligned angular momentum
in our stability calculations. This is
not only restricted to complex mergers and galaxy interactions but
may as well originate from the slow and gentle cooling of the diffuse
halo gas. In fact, simulations have shown that the present day stellar
disks in $L_*$ galaxies are built predominantly by the late
cooling of the {\it hot} halo gas component with well-aligned angular
momentum \citep{Sales2012}, whereas misalignments will tend to build the bulges and
dispersion-dominated centers of galaxies \citep{Scannapieco2009,Sales2012, Aumer2013}.
This implies that the persistence of kinematically misaligned disks is strongly dependent on the supply timescales of the (inclined) cooling gas. And therefore, that within the cosmological framework, the existence and formation of counterrotating disks and their expected timescale of survival are intrinsically connected.

In this paper we present an attempt to quantify some of these issues 
using the cosmological hydrodynamical simulation Illustris
\citep{Vogelsbergeretal2014, Vogelsberger2014b, Geneletal2014, Nelson2015}. Since we expect
misalignments to be rare, one requires large volumes explored in
projects such as Illustris, which follows the formation of tens of
thousands of galaxies with a consistent choice for the
baryonic modeling of star formation and feedback.  We focus on the
regime of low-mass (sub-$L_*$) galaxies, where most of the stellar component is built
in-situ \citep{RodriguezGomezetal2015}, which simplifies the
interpretation of the role of mergers in our results. We introduce our
sub-$L_*$ galaxy sample from the Illustris simulation in Section
\ref{sect:data}, describe the counterrotating sample in Section~\ref{sect:counter},
and discuss their origin and evolution in Section~\ref{sect:evol}. We discuss the relevant timescales in Section~\ref{sect:time}, and Section~\ref{sect:conclude} provides the main conclusions from our work.

\section{Data}
\label{sect:data}
The Illustris simulation\footnote{http://www.illustris-project.org} is a large-scale cosmological box ($106.5 \; \rm{Mpc}$ on a side) run with full hydrodynamics and galaxy formation models using the moving-meshing code {\sc arepo} \citep{Springel2010}. 
Illustris has a particle mass resolution $m_p=1.26 \times 10^6$ and $6.26 \times 10^6$ $\rm M_\odot$ for baryons and dark matter respectively, and a gravitational softening never larger than $0.7$ kpc, thereby resolving 30000 galaxies with mass $M_* \geq 10^{8.5} \; \rm M_\odot$ with at least 250 stellar particles \citep{Vogelsbergeretal2014,Vogelsberger2014b, Geneletal2014}. Halo and galaxy catalogs are built using {\sc subfind} and time evolution is studied using LHaloTree merger trees \citep{Springel2005}.  

Subgrid physics governing star formation and feedback in the simulation builds upon \citet{SpringelHernquist2003}, with the addition of stochastic winds to simulate the galactic outflows \citep[see][for details]{Vogelsberger2013}. In short, gas above a density threshold $n=0.13 \; \rm cm^{-3}$ becomes eligible for star formation and populates an effective equation of state relating temperature and pressure in an attempt to model a hot diffuse gas medium with embedded cold and dense clouds. Stars evolve following Starburst99 stellar evolution tracks \citep{Starburst95, Starburst99} and return mass, momentum and energy following stellar winds and supernova explosions. Mass, metals and tracer particles are advected with the flow following the solutions to the hydrodynamical equations on the Voronoi mesh. 

Feedback from supermassive black holes is modelled as fast and slow accretion modes.
Friend-of-friends (FOF) dark matter halos with FOF halo mass larger than $5 \times 10^{10} \; h^{-1}\, \rm M_\odot$ are seeded with a central black hole (BH) \citep{Vogelsberger2013}, that can grow in mass through mergers and accretion. After insertion, accretion onto the black hole is tracked as a fraction of the Eddington ratio, using $> 0.05$ to define the high accretion mode, implemented as a continuous thermal energy injection in the local environment, and accretion at an rate below $0.05$ of the Eddington ratio is considered in the slow mode and modelled through the injection of hot bubbles in the circumgalactic or intergalactic medium \citep{Sijackietal2007, Vogelsberger2013}. 

With such specifications, Illustris demonstrated several successes at reproducing a large, realistic population of galaxies at $z=0$ and also as a function of time, including the diversity of galaxy morphologies \citep{Vogelsbergeretal2014,Snyder2015,Rodriguez-Gomez2017}, optical properties \citep{Torrey2015}, angular momentum content \citep{Geneletal2015}, satellite colors \citep{Sales2015}, satellite metallicities \citep{Genel2016} merger rates \citep{RodriguezGomezetal2015}, and the frequency of quasar activity \citep{Sijacki2015}, among others. 

From this simulated sample, we select central galaxies, galaxies that are the most massive within their group, in the mass range $2 \times 10^9\ M_{\sun} < M_{\star} < 5 \times 10^{10}\ M_{\sun}$ (sub-$L_*$ galaxies). The motivation to explore this range is twofold: first, the low impact of mergers expected in low mass galaxies ---~which facilitates the interpretation of the results~---
and second, it is inspired by the observational data of a companion paper of dwarf galaxy kinematics (Manzano-King et al., {\it in-prep}). 

Below $M_{\star} = 2 \times 10^9\ M_{\sun}$, simulated galaxies contain $\lesssim 1000$ stellar particles and we therefore consider their structure and kinematics to be less well-resolved, possibly affecting kinematics measurements. While \citet{Penoyre2017} argue that for detailed kinematic structure a minimum resolution of $\geq 20000$ particles is needed, we focus on the general direction of the total angular momentum content for gas and stars (and not on the structural details of the resulting morphology), allowing us to explore slightly lower-mass systems. Our sample therefore comprises 11955 {\it central} galaxies at $z=0$ within our sub-$L_*$ galaxy mass range.

\begin{figure}
\includegraphics[width = 0.5\textwidth]{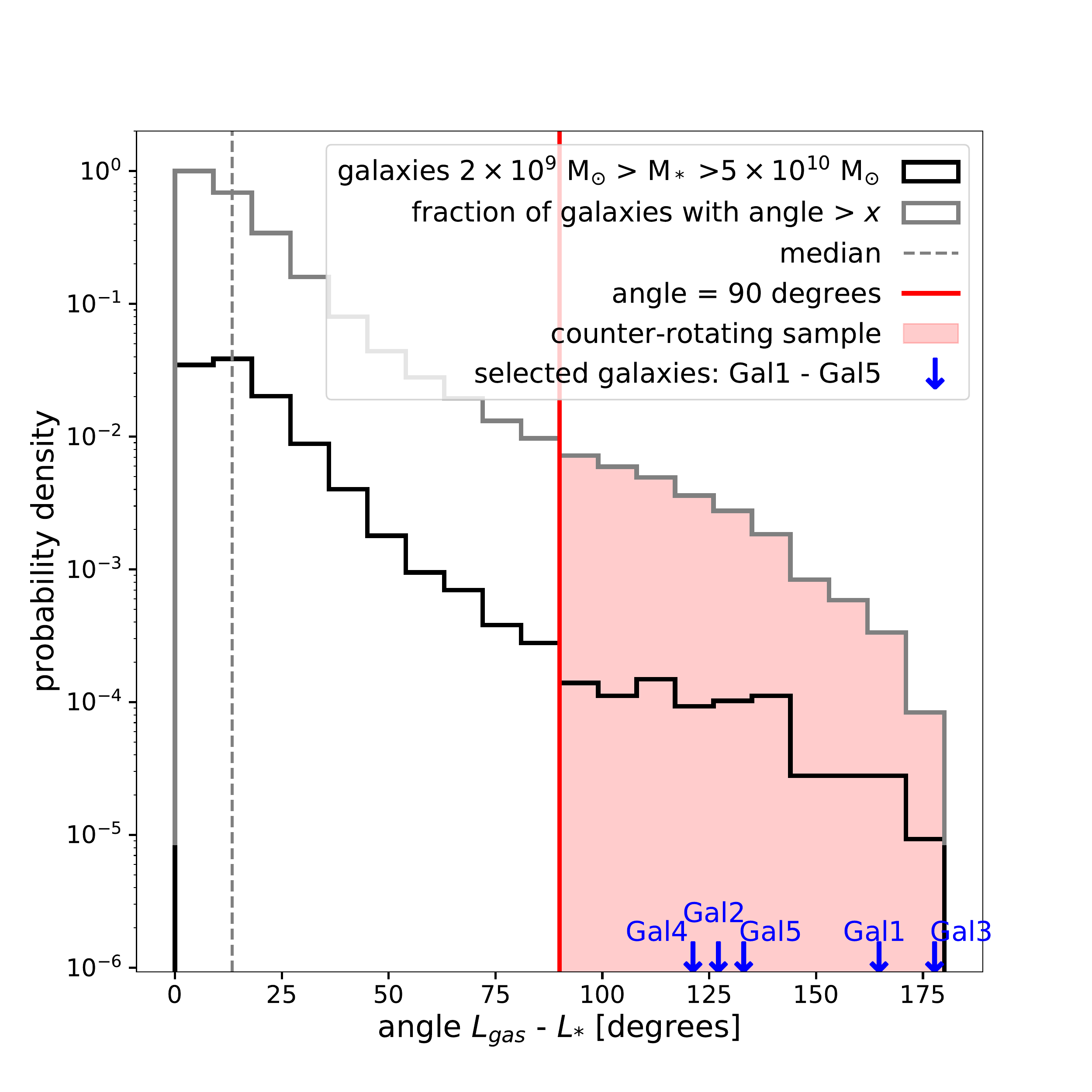}
\caption{\label{Fig:dist1} The distribution of the angle between the directions of the gas and stellar angular momentum vectors for all galaxies in Illustris with $2 \times 10^9\ M_{\sun} \leq M_{\star} \leq 5 \times 10^{10}\ M_{\sun}$ (black), and the cumulative fraction of galaxies with misalignment angle > $x$ (grey). The shaded area identifies the counterrotating systems with angles $> 90$ degrees, while the gray dashed line is the median of the whole sample ($13^\circ$). The blue arrows show the relative gas--star total angular momentum angles of the 5 galaxies described in this paper (Galaxy~1 -- Galaxy~5).} 
\end{figure}

Following previous Illustris papers, we define the galactic radius $r_{\rm gal}$ as twice the stellar half-mass radius $R_{h,*}$, and quantify galaxy properties such as mass, angular momentum, etc., using all particles within $r_{\rm gal}$. Additionally, halo gas is defined as all the gas within the subhalo at radii larger than $r_{\rm gal}$. For the dark matter instead, we will refer to quantities within the virial radius, unless otherwise specified. The virial radius is defined as the radius within which the average density is $200$ times the critical density of the Universe. To begin our analysis, we measure the angle between the total angular momentum of the stars ($L_*$) and gas ($L_{\rm gas}$) for galaxies in our sample.  The distribution of such angles is shown in Figure~\ref{Fig:dist1}.   

As expected from n\"aive galaxy formation models described in Section~\ref{sec:intro}, most galaxies tend to display a stellar and gas component that remains well-aligned within $r_{\rm gal}$. The median angle between the angular momenta for the total sample is $13$ degrees (gray dashed vertical line). However, a small number of galaxies display a large
degree of misalignment, including counterrotation between the stars and their gas. 
We indicate this by the red vertical line 
in Figure~\ref{Fig:dist1} showing $90^\circ$ misalignment between $L_*$ and $L_{\rm gas}$. With this definition, we find that 0.7\% of sub-$L_*$ galaxies in our sample display star--gas counterrotation, with the angles being preferentially in the $90^\circ$--$150^\circ$ range and only $8$ objects with an almost perfect anti-alignment ($>150^\circ$). This is surprising in view of the expectations from lifetime and stability predictions of misaligned disks. We discuss this discrepancy between predictions and numerical results further in Section~\ref{sect:time}. Small blue arrows in Figure~\ref{Fig:dist1} are used to indicate the position of Galaxy~1 through Galaxy~5, 5 randomly selected objects chosen to showcase some of our results in more detail in the following sections.

\begin{figure*}
\includegraphics[width = \textwidth]{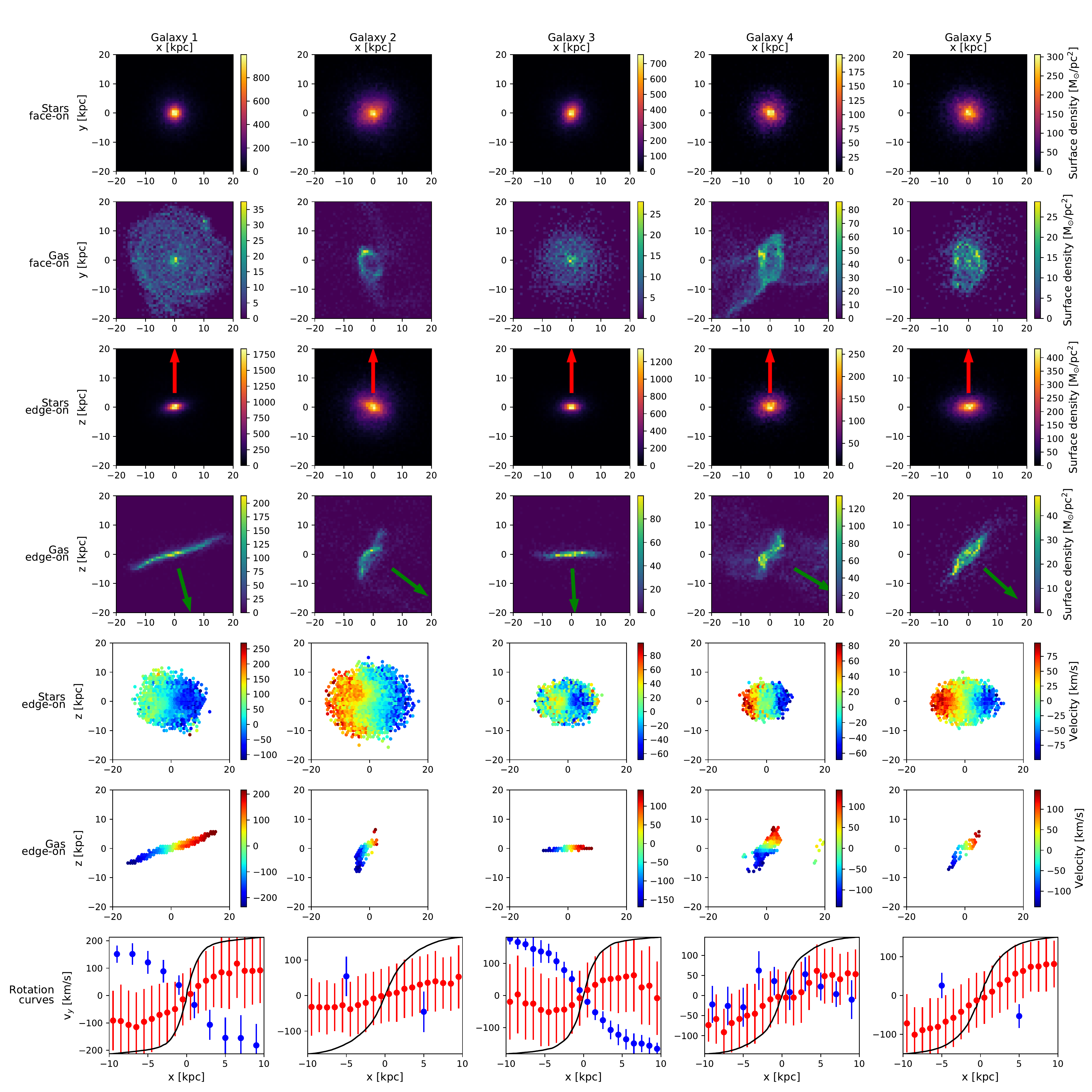}
\caption{\label{Fig:gals} The selected sub-$L_*$ galaxies with counterrotating gas, 1 per column. From top to bottom: the stellar and gas surface densities in face-on and edge-on projections, the edge-on line-of-sight velocities, and the mock slit observation rotation curves along the midplane of the stellar disk (stars in red, gas in blue, and the circular velocity plotted in black as reference).  
}
\end{figure*}

\section{Present day properties of counterrotating galaxies}
\label{sect:counter}

A visual impression of some of our objects is shown in Figure~\ref{Fig:gals}. 
Each column corresponds to one of our counterrotating galaxy examples, Galaxy~1 through Galaxy~5,
ordered according to decreasing halo mass. 
The top two rows show the face-on images for stars and gas, respectively; 
followed by the edge-on projections in rows 3 and 4. Each frame
has been rotated so that the angular momentum of the stars coincides with the $z$-axis and the total angular momentum directions of the stars and the gas are in the $xz$-plane. Red/green arrows
indicate the direction of the angular momentum for the stars and gas, highlighting the large 
level of misalignment for these 5 objects as also illustrated in Figure~\ref{Fig:dist1}. 
In both the face-on and edge-on projections it is clear that the gaseous disks 
are thinner and more extended than the stellar disks and, 
in some cases, also significantly  disturbed. 

The kinematics of these galaxies are pictured in rows 5 and~6 of Figure~\ref{Fig:gals}, which show 
the projected edge-on line-of-sight velocities for stars (row 5) and gas (row 6). 
These maps are created by considering the contribution of all stellar particles and gas cells
assigned to a given halo by {\sc subfind} and that lay within 20 kpc from its center. 
Signatures of counterrotation are evident by comparing the stellar blue/red sides (indicating velocities 
towards and away the observer respectively) which are almost exactly inverted 
for the gas (gas shows blue on the side where stars show red and vice versa).

\begin{figure*}
\includegraphics[width = \textwidth]{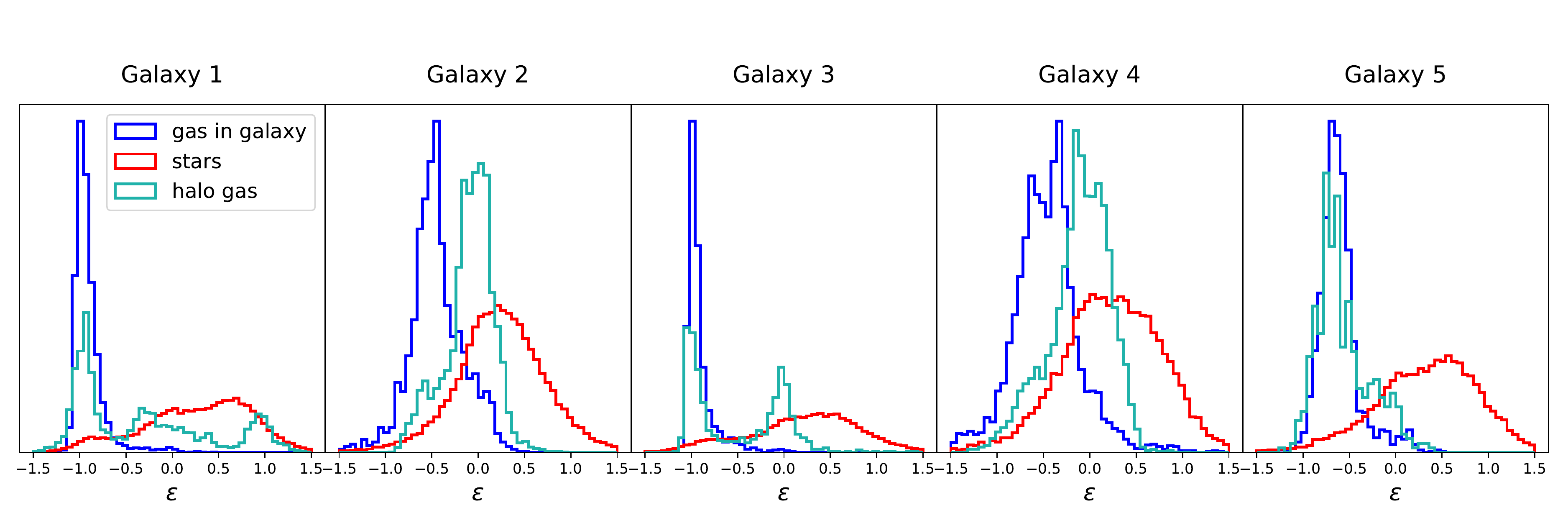}
\caption{\label{Fig:galscirc} The circularity ($\epsilon$) of the stars in the galaxy (red), gas in the galaxy (blue), and gas in the halo (light blue) for the selected sub-$L_*$ galaxies with counterrotating gas. Galaxy properties are measured within 2 stellar half-mass radii and halo gas between this and the virial radius of each object. Blue and light blue histograms extending over
negative $\epsilon$ values show that
gas within the galaxy and often also in the halo counterrotate with respect to the spin of the stars.}
\end{figure*}

The counterrotation also appears in the build up of mock rotation 
curves for these systems (row 7 in Figure~\ref{Fig:gals}).
In all cases the galaxies are "observed" edge-on with respect to the stellar disk and 
the measurements show the mean and standard deviation of the line-of-sight velocities 
in a number of bins along the midplane ($|z| < 1$ kpc) for the stars (red) and the gas (blue). 
A detailed inspection of the rotation curves in the bottom panel indicates that stars are mostly rotationally
supported but with a large degree of dispersion, whereas the gas shows in general larger circular velocities
and slightly steeper central curves. For reference, we show with a black solid line the true circular velocity
computed using $V_c^2 = GM_{\rm tot}/r$, with $M_{\rm tot}$ the total mass enclosed at a given radius $r$ in 
the simulation. As expected for these low mass systems, the maximum velocity of the gas is a 
much better tracer of the real circular velocity than the stars. Note that for inclined gas disks, the mock observed velocities along the midplane do not represent the intrinsic velocities in the inclined gas disk (see row 6 in Figure~\ref{Fig:gals}).

Alternatively, we can use the distribution of orbit circularities $\epsilon$ to scrutinize the morphology and structural
properties of gas and stars in Galaxy~1 -- Galaxy~5 (see Figure~\ref{Fig:galscirc}).
We use the definition of circularity from \citet{Abadi2003b} with the slight modifications proposed in \citet{Scannapiecoetal2009}, with $\epsilon = j/j_{\scriptsize{\rm circ}} $, where $j$ is the specific angular momentum of the particle or cell, and $j_{\scriptsize{\rm circ}} = r v_{\scriptsize{\rm circ}} = r \sqrt{GM/r}$, the expected specific angular momentum of a particle or cell at that position. 
With these definitions, large $\epsilon$ values correspond to a large degree of orbit circularity (either positive or negative)
and are typically associated with disk components, whereas bulges and dispersion dominated systems show a peak around $\epsilon \sim 0$. 

The distributions in Figure~\ref{Fig:galscirc} show a very clear separation of star particles (red) having predominantly 
positive circularity while the gas cells in galaxies (blue) show a (sometimes very sharp)
distribution around (large) negative circularity and thus a specific angular momentum content that is 
significant (peak of the distribution is around $\epsilon \sim -1$) but in the opposite direction of rotation than the stars. 
The lack of a very clear peak for the stars near $\epsilon \sim 1$ is consistent with only a moderate amount of
rotational support for these systems from the stellar rotation curves shown in the bottom row of Figure~\ref{Fig:gals}. 

For completeness, we also
show in Figure~\ref{Fig:galscirc} the orbital circularities of the halo gas (light blue histogram), 
defined as the gas beyond $r_{\rm gal}$ and within the virial
radius of each object. Interestingly, the structure of halo gas is highly complex, 
with evidence of a component in coherent motion with the galaxy gas (and therefore 
in counterrotation with the stars, by definition), but also showing a secondary peak 
around $\epsilon=0$ (all 5 Galaxies) and even a third peak
coincident with the stellar co-rotating disk (Galaxy~1).

Furthermore, note that some of our
objects, in particular Galaxy~1,~3, and~5, show a secondary excess of stars with $\epsilon \simeq -1$ and coincident with the gas component, indicating that 
---~at least in some of these galaxies~--- enough time has passed for a new generation of stars to form 
in counterrotation with the preexisting stellar disk. Examples of cases with stellar-stellar disk counterrotation
have been observed in $L_*$ galaxies \citep[e.g.][]{Rubinetal1992, Rix1992,Vergani2007,Coccato2011} and our simulations 
predict that such phenomena should also extend to lower mass systems. 

\begin{figure*}
\includegraphics[width = \textwidth]{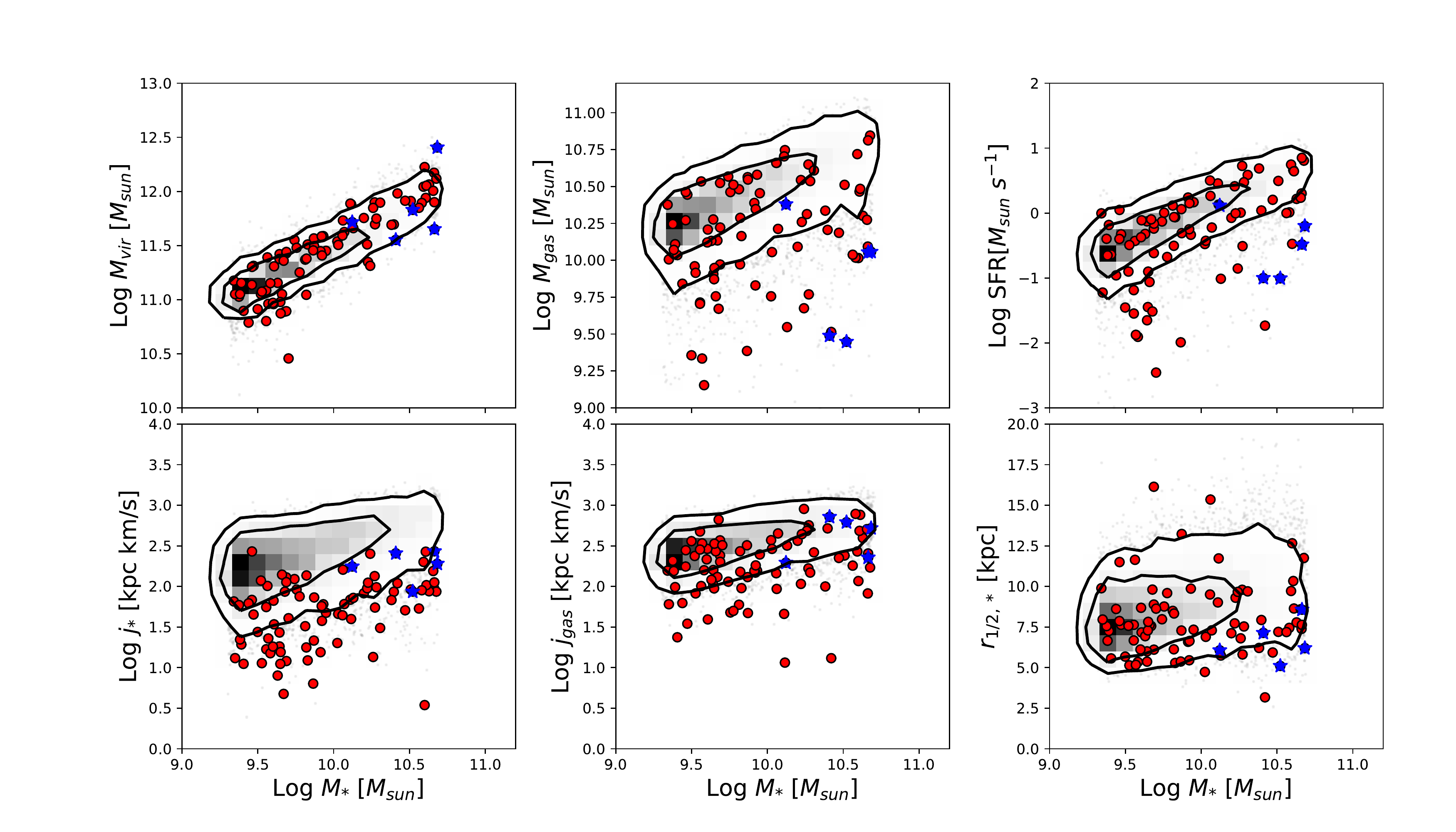}
\caption{\label{Fig:props} The distribution of galaxy properties for all galaxies within our selected mass range (grey), and the galaxies that have gas--star counterrotation (red). From left to right and top to bottom: virial mass $M_{\scriptsize{\rm vir}}$, total gas mass $M_{\scriptsize{\rm gas}}$, star formation rate (SFR), specific stellar angular momentum $j_{\star}$, specific gas angular momentum $j_{\scriptsize{\rm gas}}$, and stellar half-mass radius $R_{1/2, \star}$, as a function of total stellar mass $M_{\star}$. Galaxy~1 to~5 are highlighted (blue stars). Counterrotating galaxies have on average lower gas content, smaller
stellar and gas spins accompanied also by small stellar half-mass radii.
There is no correlation with counterrotating angle for any of these global galaxy properties.
}
\end{figure*}

After a detailed analysis of the properties of these $5$ examples, we turn our attention to the sample
of counterrotating galaxies as a whole and what their present-day properties might tell us about their formation mechanisms. 
Figure~\ref{Fig:props} shows several quantities as a function of stellar mass for the whole sample (gray contours) and 
the counterrotating objects (red circles) defined as those where the angular momentum of the gas and stars in the galaxy differ by more than $90^\circ$ (see Figure~\ref{Fig:dist1}). Galaxy~1 -- Galaxy~5 are highlighted with blue starred symbols for guidance. 

The top row of this figure indicates
that although the stellar content at fixed virial mass (top left) of counterrotating galaxies follows a similar distribution as
the whole sample, the gas content (top middle) 
at fixed $M_*$ is biased low with respect to the contours of the full population. Interestingly, the segregation 
is less clear for the star formation rate (top right), with the counterrotating sample showing on average similar star
formation levels as the control sample with similar stellar mass. 
We conclude therefore that the formation mechanism 
for these galaxies seems linked most strongly to a smaller than average gas content,
a subject we return to in Section~\ref{sect:evol}.

The bottom panels of Figure~\ref{Fig:props} show the specific angular momentum of the stars (left) and the gas (middle). 
For both quantities, counterrotating galaxies show a smaller spin than the rest of the sample ---~and in particular
for the stellar component~--- which can easily be understood
as the result of angular momentum cancellation due to the explicit counterrotation condition imposed at $z=0$ when
defining the sample, and the absence of recently formed corotating stars, in particular with large angular momenta (as would otherwise form from a large corotating gas disk). This effect would be even stronger in observed flux-weighted $v/\sigma$ measurements. Because of this low angular momentum for the stellar component, the stellar sizes tend to be on the lower end of the size distribution (bottom right panel), in agreement with analytical expectations \citep{Moetal1998} and numerical
simulations. 

In fact, \citet{Sales2012} find a fundamental link between the alignment of the angular momentum  
distribution of the baryons early in the formation time of the halo and the present-day morphology of galaxies, 
in the sense that very strong alignment is required for the formation of disks, whereas misalignments 
help build the bulges and dispersion-dominated components even in the absence of mergers. 
In our low mass Illustris sample we find a similar trend, where galaxies selected to have a counterrotating 
gas component at $z=0$ show a (stellar) morphology distribution skewed towards spheroid-dominated 
compared to the general sample in the same mass range. 

\begin{figure}
\includegraphics[width = 0.48\textwidth]{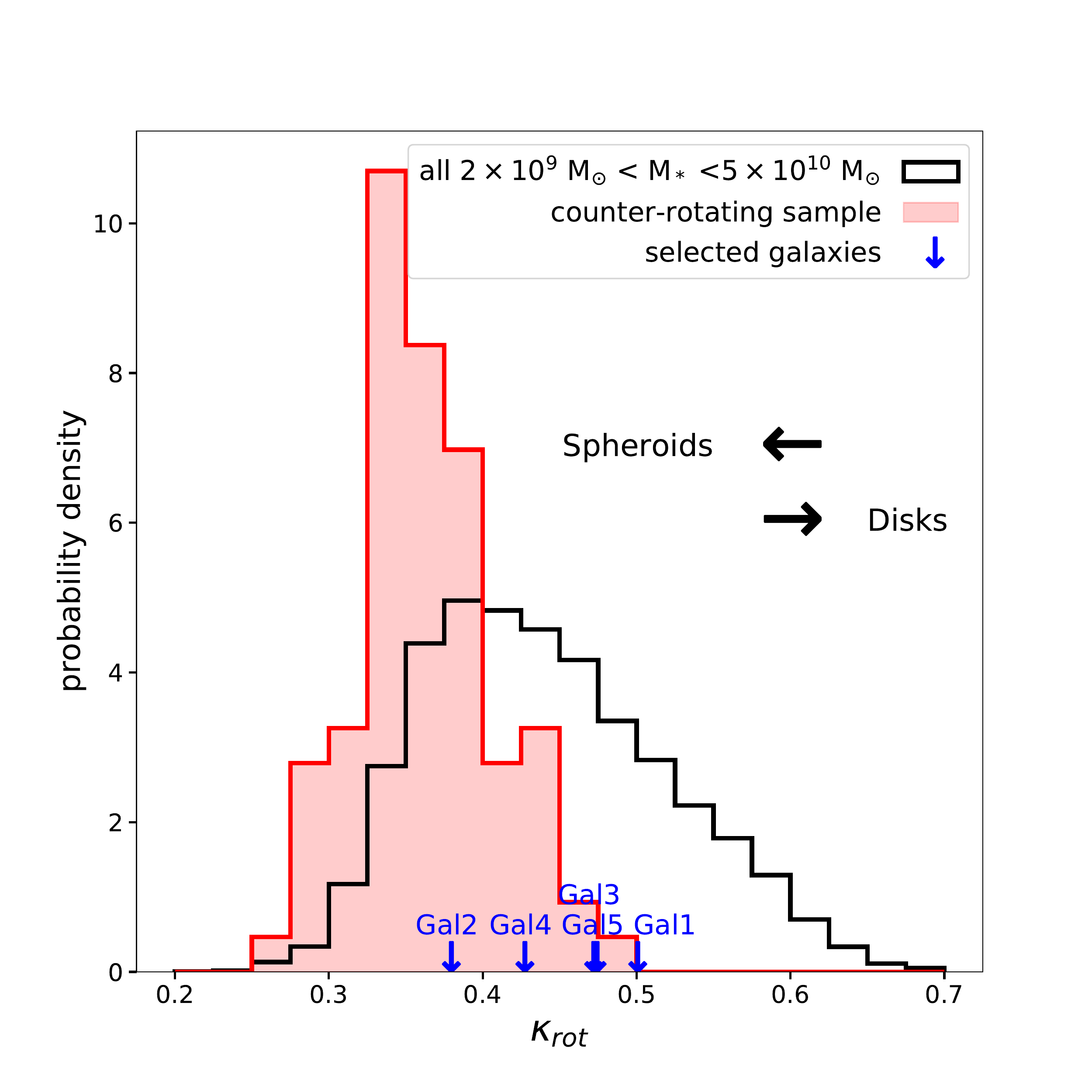}
\caption{\label{Fig:kappa} The distribution of $\kappa_{\rm rot}$ for all galaxies in our sample (black), and for the counterrotating subsample (red), with Galaxy~1 -- Galaxy~5 indicated with blue arrows. Smaller $\kappa_{\rm rot}$ correspond to more dispersion-dominated systems, like spheroidal galaxies, while large $\kappa_{\rm rot}$ corresponds to rotation-dominated systems, like disk galaxies. Gas--star counterrotating dwarfs show a stellar morphology skewed towards more dispersion
dominated objects than the full sample.}
\end{figure}

This is shown in Figure~\ref{Fig:kappa}, using the kinematics morphology indicator 
$\kappa_{\rm rot}$, defined as the fraction of the kinetic energy of the stars within a galaxy that
is in ordered rotation around the z-axis once the system has been rotated such that the total angular momentum
of the stars coincides with the z-axis \citep{Sales2010}. The solid black histogram shows all galaxies 
in our sub-$L_*$ mass range and the red shaded histogram corresponds to the counterrotating sample, with the latter 
displaying on average smaller $\kappa_{\rm rot}$ values. 
The counterrotating galaxies show lower $\kappa_{\rm rot}$ values at all masses over our selected mass range, and the variation of $\kappa_{\rm rot}$ with stellar mass for the parent dwarf galaxy sample is ${\sim}0.05$--$0.1$ from $M_* = 2 \times 10^9\; M_{\sun}$ to $M_* = 5 \times 10^{10}\; M_{\sun}$.
 
Our results therefore support the scenario where
the accretion of gas with misaligned angular momentum is a viable channel for the formation and growth
of bulges and spheroidal components in general. The correlation between alignment of angular momentum
and morphology has been reproduced by other simulations \citep[e.g., ][]{Aumer2013,GarrisonKimmel2018} 
as well as implemented in semi-analytical models
\citep{Padilla2014,Lagos2015} to follow the formation of bulges and early type galaxies.

\begin{figure}
\includegraphics[width = 0.5\textwidth]{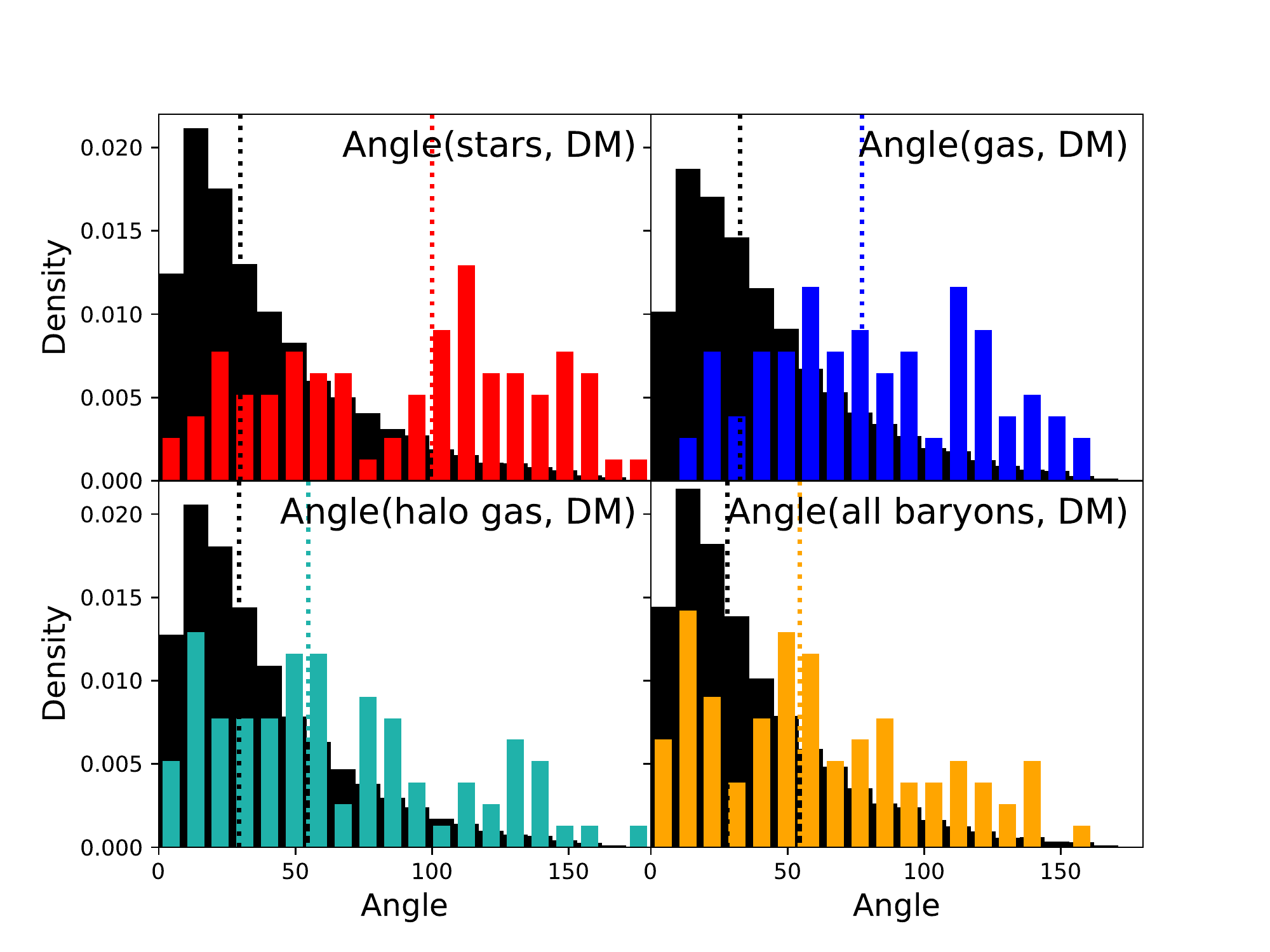}
\caption{\label{Fig:angles} Distributions of total angular momentum relative angle between the stellar disk and the dark matter halo (top left), the gas in the galaxy and the dark matter halo (top right), the gas in the halo and the dark matter halo (bottom left), and all baryons with the dark matter halo (bottom right). The dark matter total angular momentum (the total baryonic angular momentum) is calculated using all dark matter (baryonic) particles within $r_{\rm vir}$. Black histograms shows the distributions for the full sample, while colored histograms show the distribution for the counterrotating subsample. Dotted lines indicate the median values for both samples. Counterrotating dwarfs show in general stronger misalignment in all components.}
\end{figure}

A more general census of the alignments of the different components in our sample can be found in Figure~\ref{Fig:angles}. 
As before, black histograms refer to all centrals in the sub-$L_*$ sample and the colored distributions 
indicate the counterrotating objects. Each panel shows: the angle between the angular momentum of the stars in the galaxy
and the dark matter halo (top left), gas in the galaxy and the dark matter halo (top right), and 
the gas in the halo and the dark matter (bottom left). 
All components in the full sample are always better aligned with the dark matter halo than the star--gas counterrotation subsample, as indicated by the vertical lines in Figure~\ref{Fig:angles} that show the median of each distribution. This is most strongly the case for the stellar component. 

We find that
on average the stars in galaxies are aligned within $30^\circ$ with the dark matter halo whereas
for the counterrotating sample the median is $100^\circ$ (bear in mind that this sample is selected based
on gas counterrotation and not stars, so this result is not by construction). 
The bottom right panel refers to the angle between the spins of all the baryons within the virial
radius and the dark matter halo, which is similar to the halo-gas component. This is easily understood
since most of the angular momentum content is stored in the material furthest away from the galaxy due
to the linear dependence of angular momentum on distance. 
Interestingly, the gas component in the counter-rotating sample (halo gas included) is still misaligned with the dark matter, suggesting that for a significant fraction of the counterrotating 
galaxies, whatever process caused the misalignment with the stars is not affecting the angular 
momentum of the dark matter halo in the same way.

\section{The origin of strong counterrotation}
\label{sect:evol}

Gas--star counterrotation in galaxies has mostly been attributed to an external 
origin\footnote{Proposed methods for internal angular momentum transfer 
\citep{DeRijckeetal2004} and separatrix crossing \citep{EvansCollett1994} 
offer formation mechanisms for counterrotating stellar disks.} such as 
 mergers \citep[e.g.][]{PuerariPfenniger2001, Crockeretal2009,Gehaetal2005} 
and cosmological gas accretion that changes its spin orientation over time. 
Several scenarios have been identified that can lead to changes in the gas 
accretion direction, including changes in the dominant 
filament feeding the gas \citep[e.g][]{ThakarRyden1996, Bekki1998, Pizzellaetal2004, Algorryetal2014}, 
the probably related spontaneous ``spin flips" in the dark matter halo \citep{Bett2012} 
or changes in the orientation of halo shapes with radius \citep{Bailin2005}.

To disentangle the formation mechanism of gas--star counterrotating galaxies, we make
use of the merger trees in the Illustris simulations, tracing backwards in time the
counterrotating candidates at $z=0$. We explore the role of mergers on the buildup of
all of our counterrotating systems and find that none of them have had a significant merger 
(with merger ratios > 1 : 10 in total stellar mass) more recently than $\lesssim 10$ Gyr ago.
 This is not unusual for the low mass range explored here, since the overall 
 merger rate for $M_*{\sim} 3 \times 10^{10} M_{\sun}$ galaxies considering merger ratios 
 $> 1/10$ is only ${\sim}0.05\ \textrm{Gyr}^{-1}$ \citep[see][Fig.~7]{RodriguezGomezetal2015}.
We therefore conclude that mergers cannot explain the bulk of the misaligned components
in our galaxies.

\begin{figure*}
\includegraphics[width=0.49\textwidth]{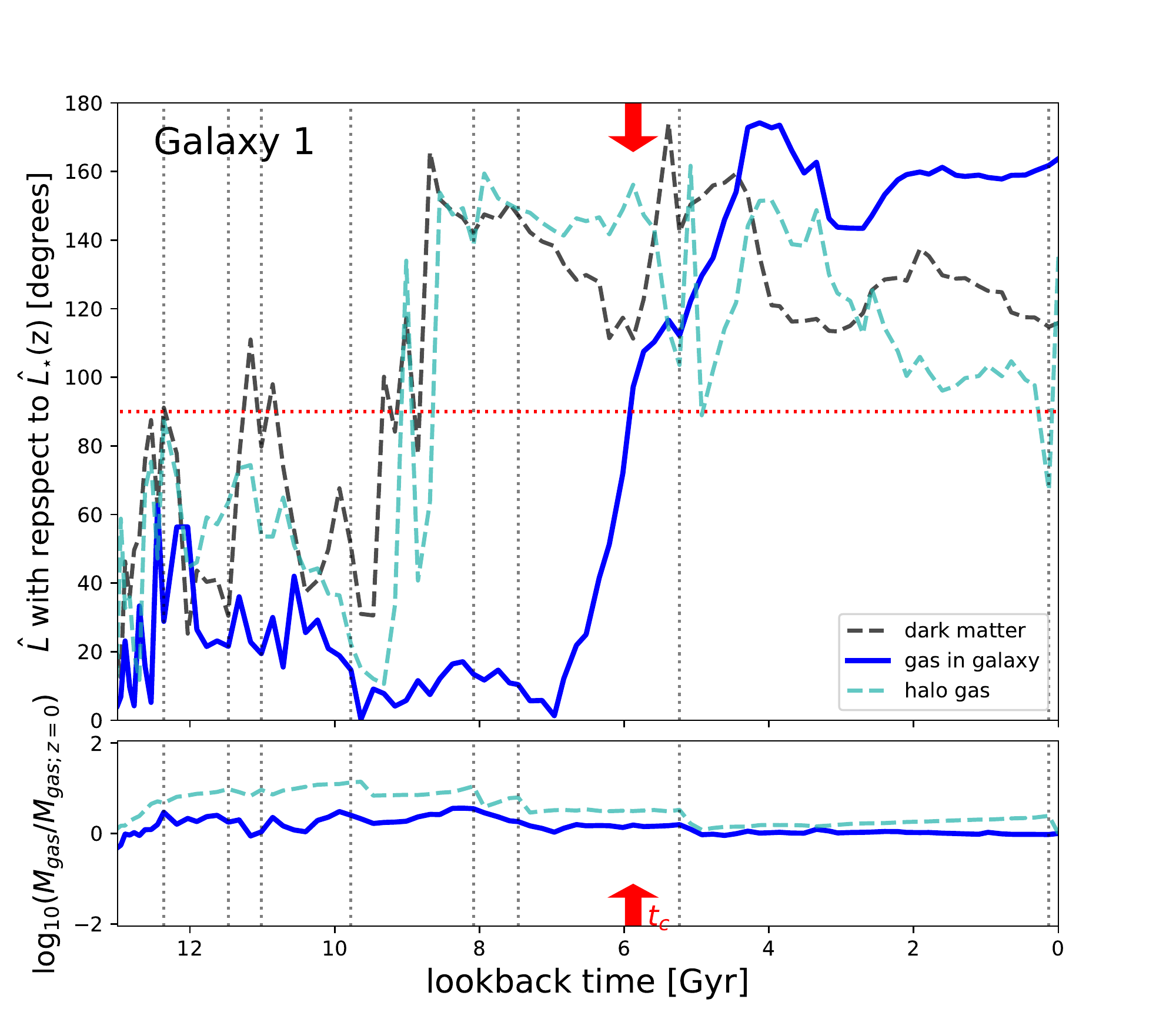}
\includegraphics[width=0.49\textwidth]{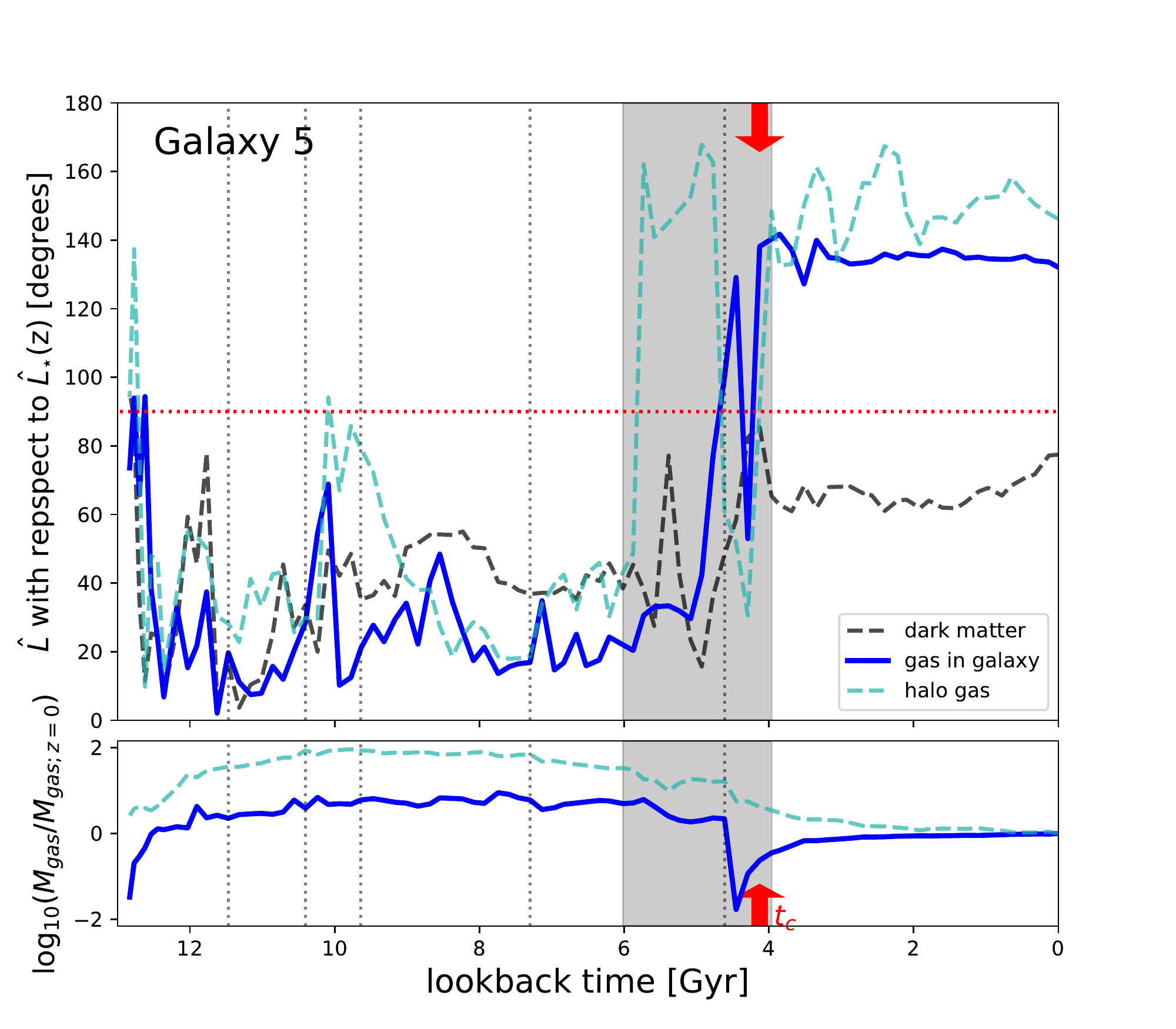}
\caption{\label{Fig:anglestime} Top panel: The evolution of the direction of the total angular momentum of the gas within the galaxy (blue), the gas in the halo of the galaxy (dashed light blue), and the dark matter halo (dashed black), with respect to the direction of the total angular momentum vector for the stellar disk at each point in time for Galaxy~1 (left) and Galaxy~5 (right). Bottom panel: The evolution of the galaxy (blue) and halo (dashed light blue) gas mass with respect to the value at $z=0$ for Galaxy~1. We highlight the two main mechanisms driving counterrotation in our sample: in both panels, light grey bands denote the periods of time that a galaxy is classified as a satellite of a larger halo, and vertical black dashed lines indicate the times a  galaxy has just experienced the slow-accretion mode of feedback from the central black hole. A galaxy component is considered counterrotating when the relative angle $> 90$ degrees (dotted red line), with the most recent change from a corotating gaseous disk to a counterrotating gaseous disk indicated (red arrows). 
}
\end{figure*}

Instead, we identify a common factor in our counterrotating galaxies: 
a significant drop in their gas mass content at some point in their 
evolution followed by the accretion of fresh gas
with misaligned angular momentum compared to the prevailing galaxy. 
We illustrate this in detail for two of our example galaxies introduced in Section~\ref{sect:counter}.
Figure~\ref{Fig:anglestime} shows the alignment history of the angular momentum of the 
star and gas within the galaxy ($L_*$ and $L_{\rm gas}$ respectively) as a function of time 
for Galaxy~1 (left) and Galaxy~5 (right).
Thick blue lines show that initially gas and stars are relatively well-aligned in both examples
until a sudden change in the orientation of $L_{\rm gas}$ causes the curves to jump to the
counterrotating regime, shown here by the horizontal red dotted line indicating relative angles 
larger than $90^\circ$. For Galaxy~1 and Galaxy~5 such a transition occurs about $6$ and $4$ Gyr ago, 
with the gas remaining counterrotating thereafter. 

A closer inspection of Figure~\ref{Fig:anglestime} reveals that the rapid transition of the 
gas orientation begins typically associated with an episode of gas loss, 
which can be appreciated as a drop in the blue solid curves in the bottom panels 
of each figure indicating the relative fraction of gas mass in the galaxies 
as a function of time normalized to the $z=0$ content. 
The interesting question then becomes, what is driving these gas mass losses in our sample? 

Two main processes are responsible for the gas removal in our counterrotating galaxies:
$(i)$ feedback from the central black hole during the slow accretion mode 
and $(ii)$ temporary ``fly-by" events through more massive systems such as groups or clusters. 
For Galaxy~1 and Galaxy~5 we indicate both of these kind of events. With black dashed vertical lines 
we highlight the 
times of injection deposition of energy due to the slow accretion mode of the central
black hole ("radio mode" feedback is activated with accretion Eddington ratio $< 0.05$).
On the other hand, using gray shading we indicate the time periods when galaxies 
are not centrals of their own halos but instead become satellites of a larger system. 
While for Galaxy~1 the change of spin orientation for the gas seems triggered by a 
feedback event at ${\sim}7.5$ Gyr ago, Galaxy~5 is a combination of both mechanisms 
acting at the same time a bit less than $5$~Gyr ago. 

Naturally, the removal of the gas is not enough to cause counterrotation, 
but it should be followed by the reaccretion of some external gas with different angular momentum. 
We note in Figure~\ref{Fig:anglestime} that in both cases the halo gas (shown as light blue dashed lines) had already acquired a counterrotating spin preceding the
mass ejection of cold gas from the galaxy. It is the subsequent partial cooling of this 
halo gas component with a misaligned angular momentum that 
defines the gas--star present day counterrotating nature of these systems. 

We examine also the relation to the spin of the dark matter halo in Figure~\ref{Fig:anglestime},
with relative angles between $L_{\rm DM}$ and $L_*$ shown in black dashed lines. 
We find that whereas for Galaxy~1 the change in spin of the halo gas is associated with the
dark matter halo change (as expected from tidal torque theory arguments), 
in Galaxy~5 the same does not hold, with the dark matter spin remaining relatively stable 
with respect to the stars in ${\sim}30$--$70^\circ$ at all times. However, we note that for Galaxy~5
the change of halo gas spin is environmentally induced. This object
passes through the outskirts ($r \geq 300$~kpc) of a system ${\sim}100$ times more massive
during which drag forces and interactions with the intra-group medium changes the angular momentum 
direction of the halo gas while the galaxy is temporarily a satellite (gray shaded area).
These two examples also nicely illustrate the complexity of angular momentum acquisition
in baryons once the full cosmological assembly of galaxies is considered.

\begin{figure}
\includegraphics[width = 0.5\textwidth]{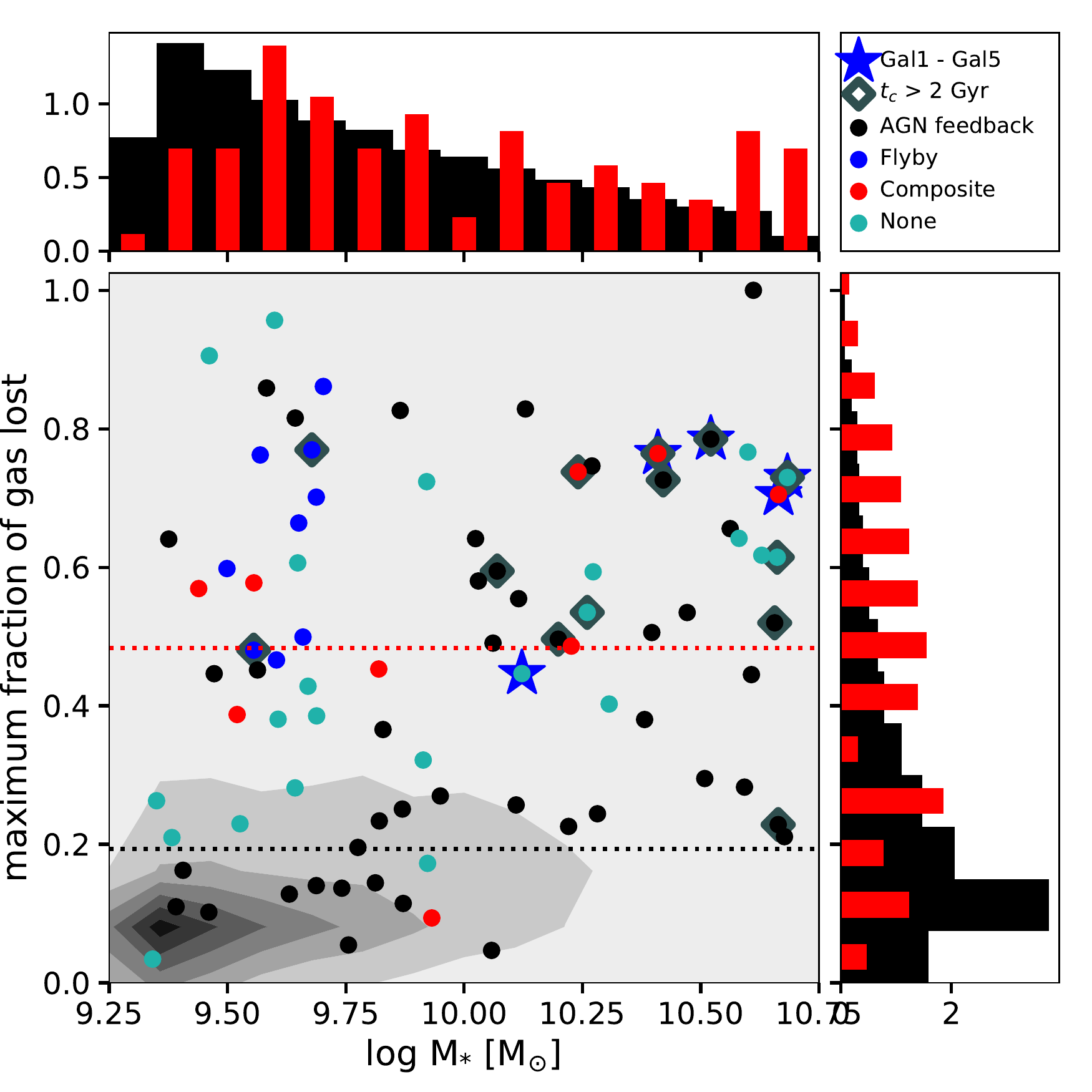}
\caption{\label{Fig:gasloss} The distribution of the maximum amount of total gas mass lost to the subhalo with total present-day stellar mass for all central galaxies at $z=0$ in the Illustris simulation within our mass range $2 \times 10^9\ < M_{\star}/M_{\sun} < 5 \times 10^{10}$ (gray background and black histograms), and for the subset of present-day centrals that have counterrotating gaseous disks (scatter points and histograms). Galaxy~1 -- Galaxy~5 are highlighted with blue stars, and galaxies that have been counterrotating for longer than $2$~Gyr are highlighted with gray diamonds. All counterrotating galaxies are color-coded by their possible cause for counterrotation: AGN feedback (black), flyby interactions (blue), a combination of both (red), or neither (light blue). Horizontal dotted lines indicate that the median gas mass removal in the full sample is about~$20\%$, which is lower than the typical episode for our counterrotating sample, with about~$50\%$ of the gas removed in a single episode.}
\end{figure}

The other 3 galaxies studied in detail in Figure~\ref{Fig:gals} show a similar combination
of the effects described above and we include their equivalent alignment evolution in Appendix~\ref{app}
for completeness. 

We now generalize the argument of the link between gas--star misalignment and gas loss 
events to the whole counterrotating sample in Figure~\ref{Fig:gasloss}. We show as a function of 
$M_*$ the maximum fraction of gas lost within the galaxies during their time evolution. The gas loss is calculated as the fractional difference between the beginning and end of any period of continuing decrease in gas mass. 
While for the whole population the median gas mass variation is under~$20\%$, for our counterrotating
sample (colored symbols) the median gas fraction lost is about half the gas content in a single
episode. This is particularly true for objects where counterrotation has settled more than $2$~Gyrs ago (larger symbols).
We therefore conclude that significant gas removal is a general feature of our counterrotating 
sample and does not apply exclusively to the specific cases of Galaxy~1 and~5 showcased in Figure~\ref{Fig:anglestime}. 

Moreover, we confirmed the prevalence of the two mechanisms of gas removal described above
for Galaxy~1 and Galaxy~5,
namely slow mode black hole accretion feedback or temporary 
fly-by through a more massive system, for the large majority of the counterrotating sample. 
Based on the \emph{most recent} time at which counterrotation sets in $t_c$ (i.e. when the
relative angle between $L_{\rm gas}$ and $L_*$ crosses~$90^\circ$) we flagged 
each counterrotating system according to: ``black hole induced'', 
``fly-by induced'', ``composite origin'', or ``none of these'', finding that
${\sim}73\%$ of the counterrotating galaxies satisfy at least one of these criteria.

\begin{figure}
\includegraphics[width = 0.5\textwidth]{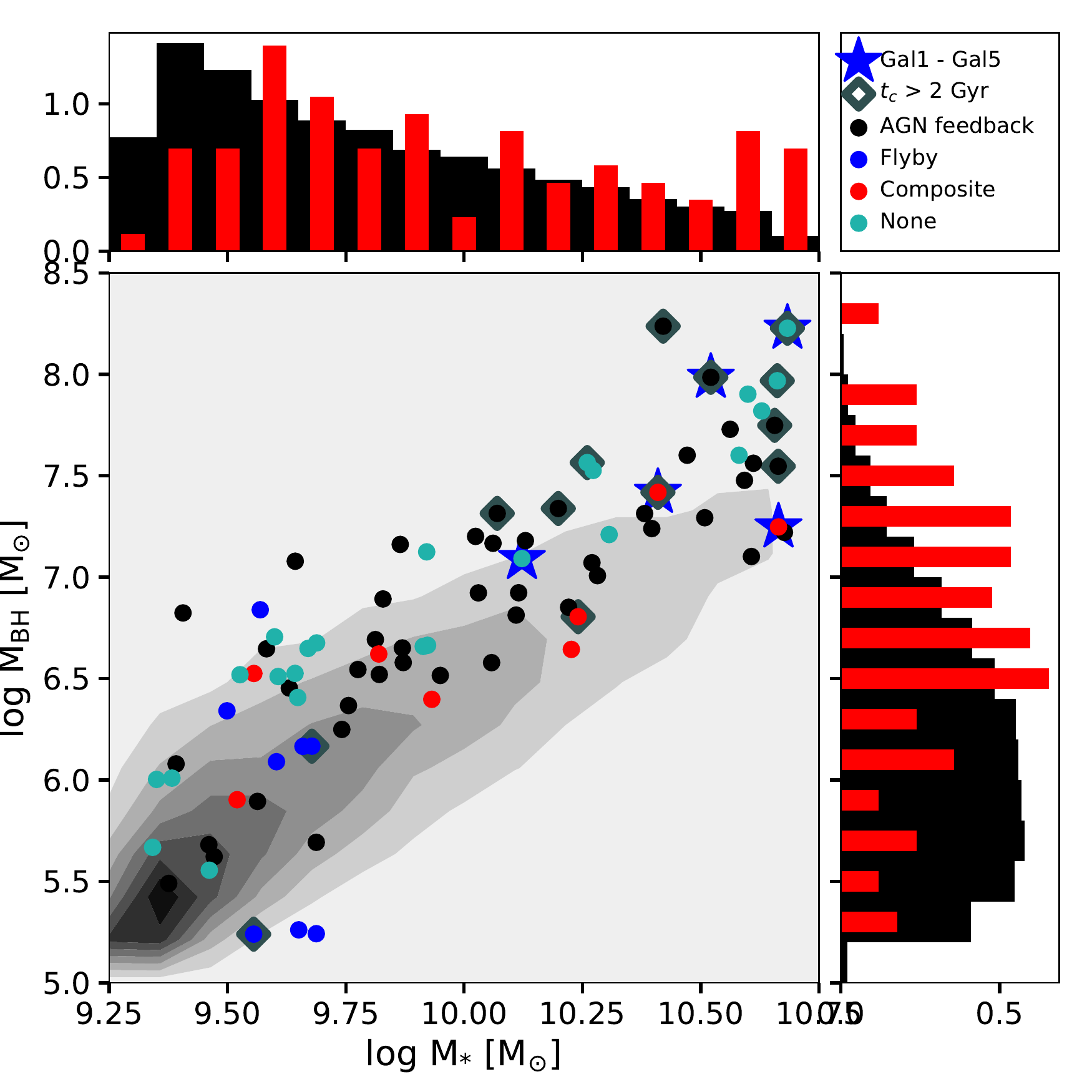}
\caption{\label{Fig:mBH} The black hole mass -- stellar mass distribution for our full sample (black), and the counterrotating subsample (scatter points and red histogram). See Figure~\ref{Fig:gasloss} for a description of the scatter point markers and colors. Counterrotating low mass galaxies have larger black hole masses at a given $M_*$, in agreement with the
role played by black hole feedback in setting the counterrotation in our sample.}
\end{figure}

For this classification we looked for temporal correlations between the
stellar-gas counterrotation and feedback/fly-by events. 
We proceeded as follows. We marked them as ``black hole induced" 
(black in Figure~\ref{Fig:gasloss}, \ref{Fig:mBH} and~\ref{Fig:time}) if the measured $t_c$ was within $1$~Gyr before
any of the times at which feedback from the slow black hole accretion mode
was active, or ``fly-by induced" (blue in Figure~\ref{Fig:gasloss}, \ref{Fig:mBH} and~\ref{Fig:time}) if $t_c$ was within the same time period of the
galaxy being a temporary satellite, or ``composite" (red in Figure~\ref{Fig:gasloss}, \ref{Fig:mBH} and~\ref{Fig:time}) if both conditions were satisfied (for
example as Galaxy~5 in Figure~\ref{Fig:anglestime}). We find that $52.3\%$ and~$10.5\%$ correspond
to black hole and fly-by induced counterrotation, respectively, while~$10.5\%$ can be attributed to a combination
of both effects. The remaining~$26.7\%$ (light blue in Figure~\ref{Fig:gasloss}, \ref{Fig:mBH} and~\ref{Fig:time}) could not be explained 
by these processes. We have also checked that in about $73\%$ of
the star--gas counterrotation episodes the halo
gas is counterrotating within the $1$~Gyr before the gas in the
galaxy counterrotates, indicating that the change in spin is
at the global scale of the halo, not only the galaxy.
In partial support of the connection to black hole feedback in our simulations, we find that
counterrotating galaxies tend to host a more massive black hole at a fixed stellar mass
compared to the population as a whole (see Figure~\ref{Fig:mBH}).

We hasten to add that results from our simulations might be bound to the specific
modeling of star formation and feedback in Illustris. 
Although it is tempting to draw a direct link between structural features such as misaligned gas disks
to black hole feedback in sub-$L_*$ galaxies, one should as well consider the possibility
that the effect is limited to the numerical choices for the implementation of this kind of
feedback in our simulations. 

This is difficult to assess. On one hand, black hole feedback in Illustris
has been found to be overly efficient in low mass groups, depleting the gas content of 
$M_{\rm vir} \sim 10^{13} \; \rm M_\odot$ groups \citep{Geneletal2014}. On the other hand, the same flavor of
feedback is not efficient enough in more massive systems to completely halt star formation, 
resulting in central cluster galaxies that are too blue and with a large fraction of in-situ
star formation \citep{Rodriguez-Gomez2017}. It is therefore difficult to predict the accuracy of the treatment of
black hole feedback for our centrals, which populate halos with smaller masses
$M_{\rm vir} \sim 10^{11}$ -- $10^{12.2} \; \rm M_\odot$ (top left panel Figure~\ref{Fig:props}), 
well within a mass range where observational constraints are scarce.

As a cautionary first step we have checked the fraction of counterrotating gas--star 
galaxies in EAGLE (\citealt{Schayeetal2015, Crain2015, McAlpine2016}; for galaxy angular momentum and misalignments see \citealt{Velliscig2015, Zavala2016, Lagos2017}) and IllustrisTNG \citep{Pillepichetal2018,Weinbergeretal2017, Nelson2018a, Pillepich2018, Springel2018, Naiman2018, Marinacci2018, Nelson2018b}, 
which both include completely different treatments of black hole feedback, finding that fractions in these simulations are higher than our estimates in Illustris ($0.7\%$ compared to IllustrisTNG: $6.9\%$ and EAGLE: $13.9\%$). Therefore, at face value
there is no reason to believe that our Illustris results are significantly over-predicted due to the particular
black hole feedback model. 

With these important caveats in mind, we conclude that 
the physical mechanism identified here through which gas--star counterrotation in galaxies 
arises as the combined result of gas removal and later re-accretion of misaligned gas 
stands valid, regardless of the details on the particular fraction of cases where this 
is prompted by a feedback event or by an environmental effect. Both identified channels 
present a novel and plausible way to form counterrotating low mass galaxies. 
Furthermore, our scenario makes two interesting predictions. 
First, counterrotation should be more common for dispersion-dominated dwarf galaxies than for disky ones. 
Second, objects displaying counterrotating components in this mass range should populate the 
upper end of the stellar mass -- black hole mass relation. Both predictions may offer a path
to observational confirmation of our results in the near future. 

\section{Timescales of counterrotation}
\label{sect:time}

\begin{figure}
\includegraphics[width = 0.5\textwidth]{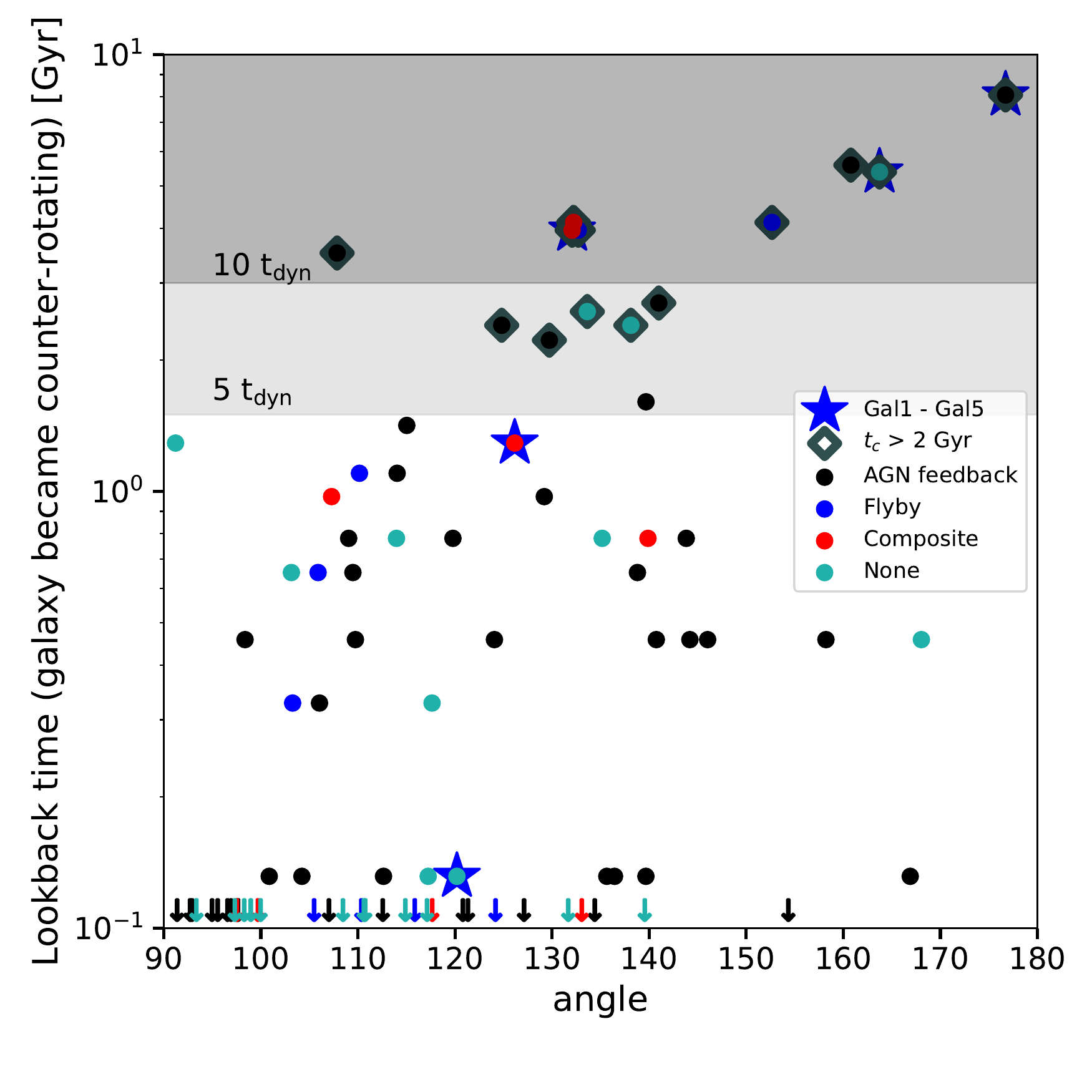}
\caption{\label{Fig:time} The relative angle between the present-day gas and stellar total angular momentum directions 
with lookback time at which they \emph{last} became counterrotating for all counterrotating galaxies. The shaded regions indicate 5~dynamical times (light gray) and 10~dynamical times (dark gray). Scatter point markers and colors are like described in Figure~\ref{Fig:gasloss}. About 25\% of our dwarfs have remained counterrotating for more than $2$ Gyr.}
\end{figure}

Finally, we turn our attention to the stability of counterrotation and the
typical timescales for which misalignments between gas and stars prevail. 
Figure~\ref{Fig:time} shows the distribution of counterrotation times $t_c$ 
introduced above, and defined as the most recent time the galaxy gas spin crossed from corotation to counterrotation, 
with an angle larger than $90^\circ$ with respect to the stellar angular 
momentum $L_*$. Surprisingly, counterrotation can be a very long-lasting
feature, with galaxies showing stable misaligned disks for several Gyr,
in particular for cases where the misalignment is large ($>140^\circ$).

As discussed in Section~\ref{sec:intro}, idealized theoretical models favor a rapid
realignment of tilted components as the combined result of differential precession 
of the disk in the non-spherical gravitational potential, and friction between concentric
gas annuli with different angular momenta that will tend to cancel out to give
a coherently rotating disk in the same plane as the stars. Because these processes
should take place quickly, a few orbits are enough to bring the system to equilibrium
requiring $2$--$5$ orbital times to erase counterrotation 
\citep[see][and references therein]{vandevoort2015}. 

Assuming a mass $10^{10} \; \rm M_\odot$, a circular velocity of $100$~km/s and 
at an average size of $5$~kpc for our galaxies, we estimate an orbital time
$t_{\rm dyn} \sim 300$~Myr. We indicate the regime where $t_c \geq 5 t_{\rm dyn}$ and where $t_c \geq 10 t_{\rm dyn}$ 
with shading in Figure~\ref{Fig:time}, finding that in the believed unstable 
angle ranges (not perfectly aligned, not $90^\circ$, and not perfectly misaligned: ${\sim}120^\circ-{\sim}160^\circ$) counterrotation can survive well past a couple
of dynamical times. Moreover, for the inner regions, dynamical times shorten 
steeply with radius, resulting in even longer relative timescales for survival of
the counterrotating component. This seems to hold regardless of the particular mechanism
giving rise to the counterrotation, as indicated by the different symbols in Figure~\ref{Fig:time}.

In agreement with previous simulations \citep{Brook2008, Roskar2010, Algorryetal2014, vandevoort2015} 
the origin for the persistence of the tilted
gas component seems to be the continuous supply of misaligned gas from the more external
regions of the halo (as shown for the Galaxy~1 -- Galaxy~5 examples in Figure~\ref{Fig:anglestime} and
Appendix). This means that once counterrotation is detected, the probabilities of finding
a larger reservoir of outer halo misaligned gas should be high. It also means that because
counterrotation can be so long-lived, there may remain no observational evidence at the present
day of the mechanism creating the counterrotation in the first place. This should be taken into
account when scrutinizing present-day environment or black hole activity in observational 
samples of these galaxies. However, the existence of counterrotation can itself be seen as a tracer of past environmental changes, or strong feedback.

\section{Summary and Conclusions}
\label{sect:conclude}
We describe the population of gas--star counterrotating low-mass galaxies in the Illustris 
simulation analyzing a sample of 11955 central galaxies in the stellar mass range 
$2 \times 10^9\ M_{\sun} < M_* < 5 \times 10^{10}\ M_{\sun}$. We define a system to 
be gas--star counterrotating when the relative angle between the total angular momentum 
vectors of all the stellar particles and gas cells within two stellar half-mass radii 
is larger than $90$~degrees. Our findings can be summarized as follows. 

\begin{itemize}
    \item The Illustris simulation produces a very low fraction of gas--star counterrotating sub-$L_*$ galaxies, $0.7\%$. 
    For these galaxies, the spin of the stars as well as that of the gas show large deviations from that of the dark matter halo, with average angles $100^\circ$ and $80^\circ$, respectively; indicating a complete decoupling from the dark matter halo spin. On the other hand, the normal population displays a stellar spin alignment on average within $30^\circ$ of that of the dark matter halo component (and similarly with the gas). 
    \item The present-day global properties of our gas--star counterrotating galaxies agree well with the control sample except for a significant mean overall ${\sim}30\%$ gas deficit compared to galaxies at fixed $M_*$, and a low angular momentum content of both the stellar and gas components. The mean star formation rate however, shows no significant deviation from the mean of the normal population. The low angular momentum 
    content of the stellar components found for the sample of counterrotating galaxies
    agrees well with previous results from simulations highlighting that angular momentum 
    misalignments help build dispersion dominated stellar components in galaxies such as 
    bulges and spheroids.
    \item We find no relation between gas--star counterrotation and the presence of merger events with mass ratios 
    larger than 1:10. Instead, most our systems have experienced in the past a significant gas removal event 
    followed by the reaccretion of new gas with a misaligned spin. 
    \item Two main mechanisms drive the gas removal originating the counterrotation in our systems: black hole 
    feedback (and in particular the slow accretion mode) and environmental effects during fly-by encounters with more massive systems. As a result galaxies that exhibit counterrotation today may hold clues to a past violent feedback episode or to a complex environmental history. 
    \item Once established, gas--star counterrotation may survive for several Gyr (with $15\%$ of our sample displaying
    formation times more than 2 Gyrs ago), in contrast to classical theoretical estimates of the quick action of 
    torques to re-align the components. We attribute this to the presence of a continuous supply of halo-gas 
    which also shows a large degree of misalignment. 
\end{itemize}

As discussed in Sec.~\ref{sect:evol}, the prevalence of black hole feedback driving these misalignments in our sample may be the result of the particular feedback modeling in Illustris. The take away point from this analysis is the link between present day gas counterrotation and a past event of gas removal that dissolves the corotating gas disk paving the way for reaccretion of misaligned new gas. Any feedback source able to efficiently couple to the interstellar medium can offer an avenue towards this. Reassuringly, we have checked that in IllustrisTNG and EAGLE (which both include completely different treatments of black hole feedback) the fraction of counterrotating systems is even larger than in this work, suggesting that our results do not depend only on an overly-efficient black hole feedback model. Furthermore, Serra et al. (in prep.) using the EAGLE simulations \citep{Schayeetal2015} find a similar connection with gas removal for counterrotation to arise in more massive galaxies. 

Therefore we conclude that counterrotation in low mass galaxies is associated with gas loss events driven either internally (feedback), by external factors (gas stripping from environment) or by a combination of the two; and may provide important clues on the past history of these galaxies. If the main driver is black hole feedback, it may be worth looking in observations for associations between AGN activity and counterrotation in low mass dwarfs, work that is currently being carried on by our team (Manzano-King et al., {\it in-prep}). \\

\begin{acknowledgements}
We thank the Illustris collaboration, the Illustris TNG collaboration, 
and the Virgo Consortium for making their simulation data publicly available. TKS and LVS thank Amina Helmi for valuable discussions at the inception of this work.
LVS and GC acknowledge partial support from HST-AR-14582 and NSF-1817233 grants 
and from the Hellman Foundation.  
This research made use of NumPy \citep{numpy}, matplotlib, \citep{matplotlib}, and the corner visualization module \citep{corner}. Some of the computations in this paper were run on the Odyssey cluster supported by the FAS Division of Science, Research Computing Group at Harvard University. The EAGLE simulations were performed using the DiRAC-2 facility at Durham, 
managed by the ICC, and the PRACE facility 
Curie based in France at TGCC, CEA, Bruy\`{e}res-le-Ch\^{a}tel. 
The Flatiron Institute is supported by the Simons Foundation. 
\end{acknowledgements}

\appendix
\counterwithin{figure}{section}
\section{The origin of counterrotation for Galaxies~2, 3, and 4}
\label{app}

\begin{figure}
\includegraphics[width=0.5\textwidth]{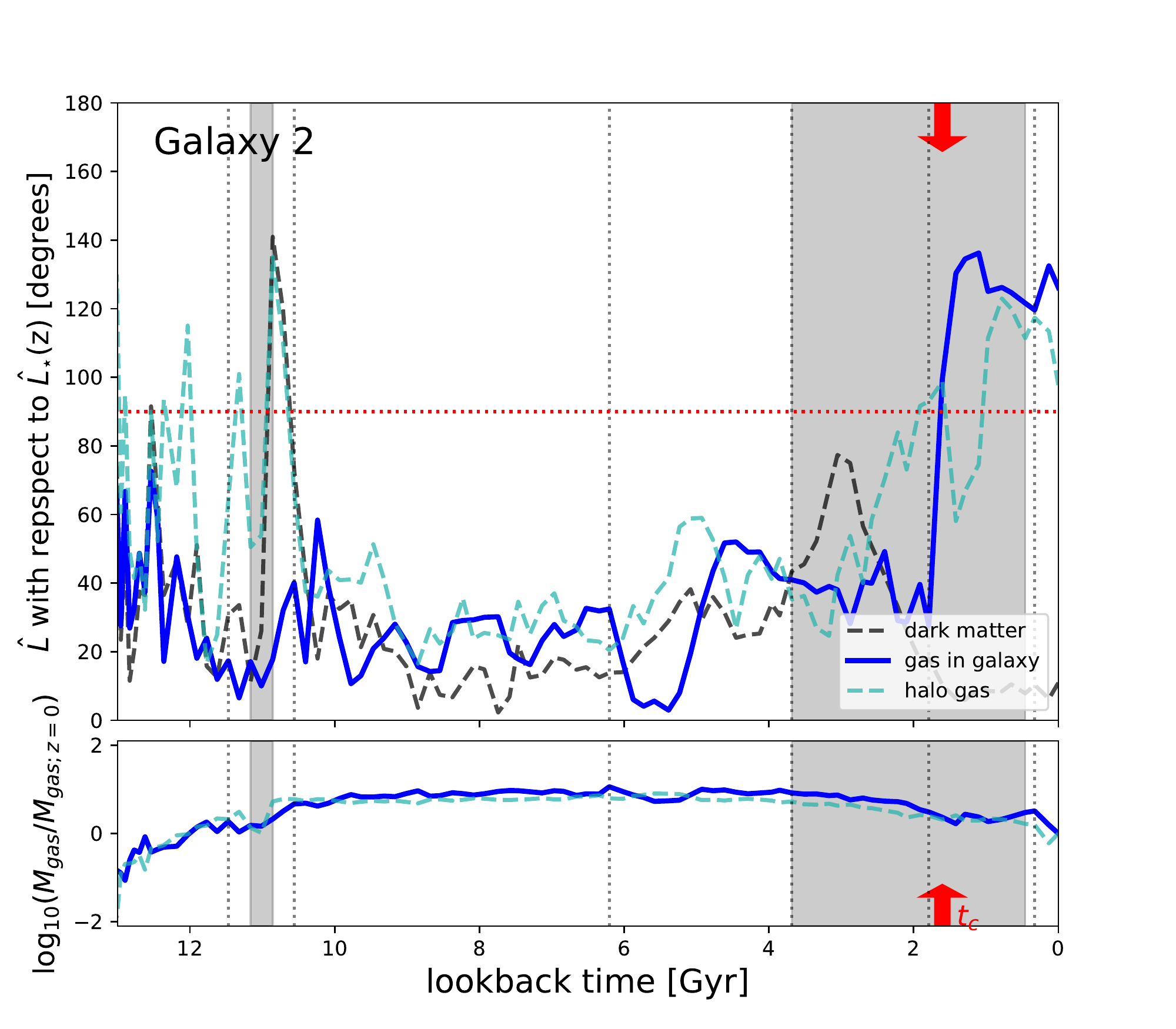}
\includegraphics[width=0.5\textwidth]{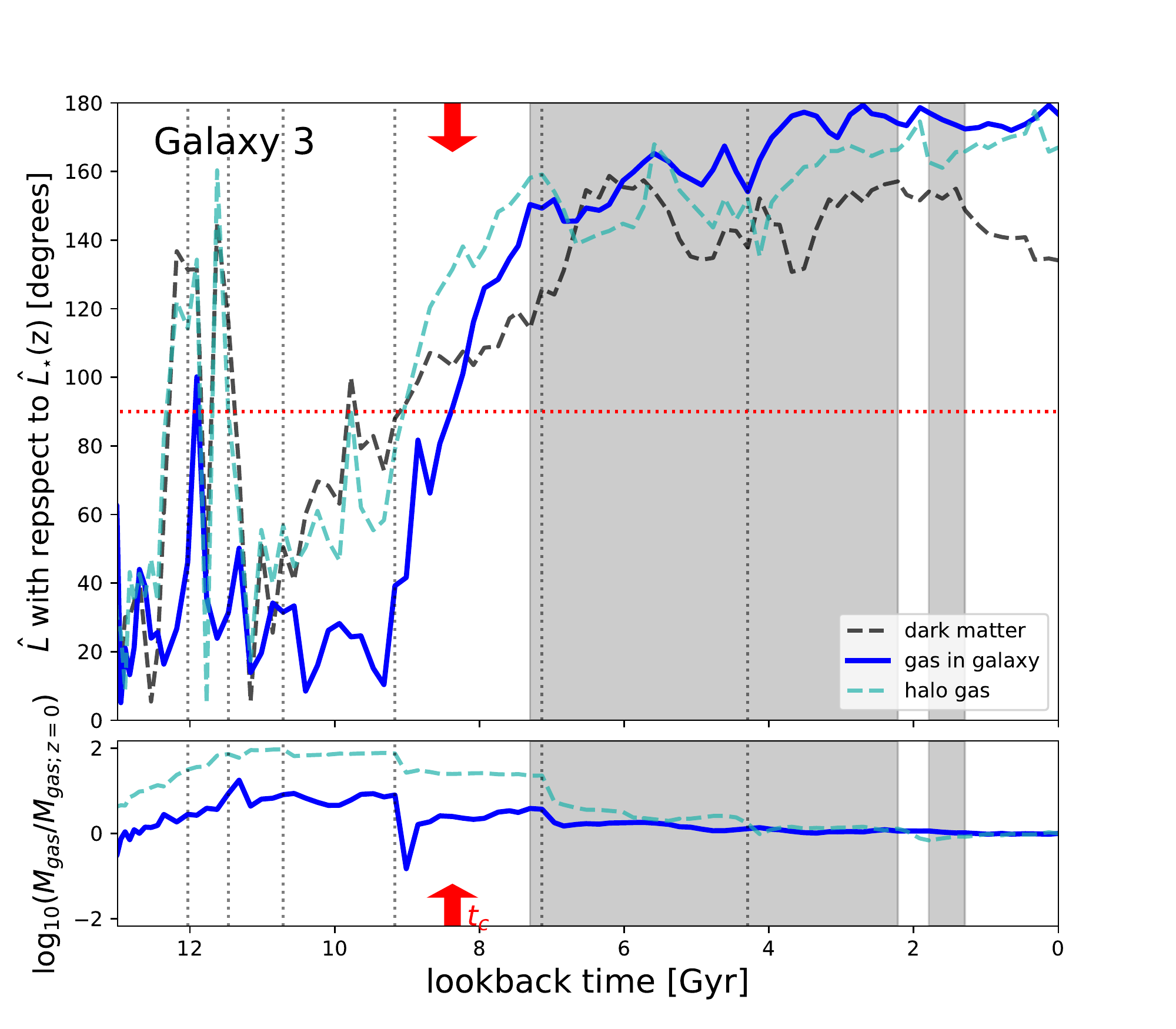}\\
\includegraphics[width=0.5\textwidth]{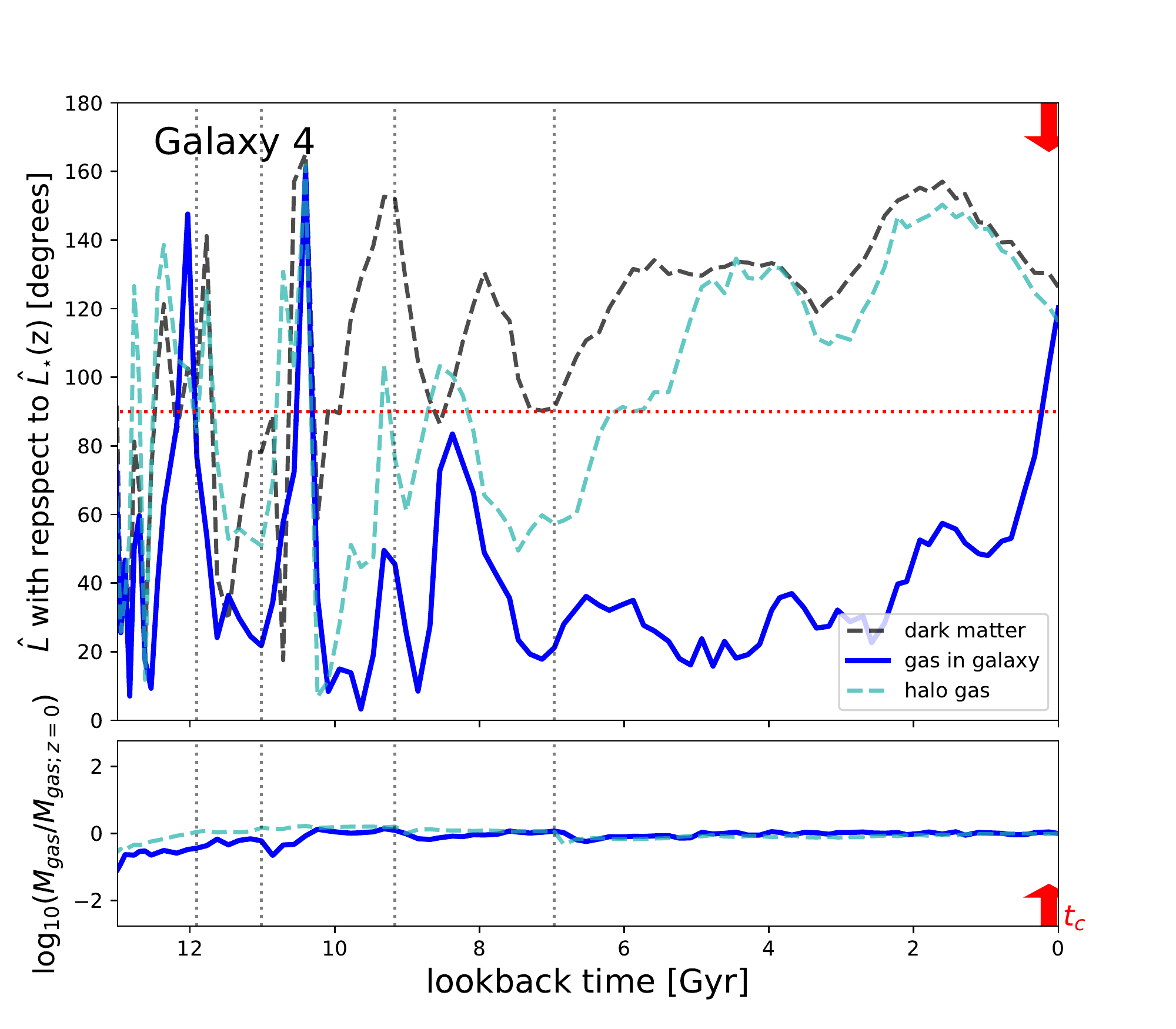}
\caption{\label{Fig:appendix1} Like Figure~\ref{Fig:anglestime}, for Galaxy~2 (top left), Galaxy~3 (top right), and Galaxy~4 (bottom left).}
\end{figure}

Figure~\ref{Fig:appendix1} shows the evolution of relative angles between the total angular momentum vector of the stellar disk and gaseous disk for Galaxy~2, 3, and~4. We will here briefly describe the origin of counterrotation for these galaxies.

Galaxy~2 shows a similar evolution as Galaxy~5 but the episode of being a satellite and the accretion of counterrotating gas takes place at more recent times. Although Galaxy~2 is ${\sim} 500\ \rm{kpc}$ away from its more massive host, its host galaxy is more than $300$ times as massive, and Galaxy~2 loses 75\% of its gas (within the galaxy, in the gaseous halo this is 61\%). The AGN feedback while the galaxy is a satellite is much less important for the gas loss in this case: it removes ${\sim} 2 \times 10^9\ M_{\sun}$ of the total $2.4 \times 10^{10}\ M_{\sun}$ that is lost due to the tidal interaction. Therefore, the environment is the key facilitator of the accretion of counterrotating material in Galaxy~2.

Completely different in evolution from the other four galaxies, Galaxy~3 has had a counterrotating gaseous disk for more than 8~Gyr. Additionally, the total angular momentum vectors of the stars and the gas in the disk or halo are almost exactly opposite. While Galaxy~3 is a satellite for a long period of time, this occurs when the counterrotating gaseous disk is already established, and while some gas is lost this is not sufficient to alter the configuration. During its early formation the orientation and angular momentum of the stars, gas, and dark matter change rapidly as many filaments feed the galaxy and the merger rate is high. Additionally Galaxy~3 experiences a strong burst of AGN feedback around a lookback time of 9~Gyr which blows out 98\% of the gas ($1.387 \times 10^{10}\ M_{\sun}$ of the $1.414 \times 10^{10}\ M_{\sun}$). The new gas accretion is dominated by gas with opposite angular momentum compared to the existing stellar disk and the new counterrotating gaseous disk grows from there. The stellar rotational velocity of Galaxy~3 in Figure~\ref{Fig:gals} has large dispersions and the circularity distribution of the stars shows a secondary peak at negative circularity (40\% of the stellar particles have $\epsilon < 0$). This suggests the existence of a secondary stellar disk that formed out of the counterrotating gas.

Galaxy~4 differs from the rest in that it has only recently become counterrotating. For most of its evolution the angular momentum of accreted material varies strongly, which results in a slightly counterrotating halo gas for the last 6~Gyr. In this case, halo gas seems to follow the spin of the dark matter halo. In those 6~Gyr that counterrotating halo gas has slowly been accreted onto the galaxy and only at the present day the gaseous disk is counterrotating. However, as shown in Figure~\ref{Fig:gals} the inner and outer gaseous disks have different rotation,
and the gas component of Galaxy~4 appears the most morphologically disturbed in Figure~\ref{Fig:gals}. Moreover, stellar rotation of Galaxy~4 is the least well-defined of the five systems, and its peak velocity is below $100\ \rm{km/s}$. The counterrotating gaseous disk in Galaxy~4 may thus be a transitory phenomenon unlike what we see in the other 4 galaxies.

\bibliographystyle{aa}
\bibliography{OnTheOriginOfStarGasCounterrotationInLowMassGalaxies}
\end{document}